\pdfoutput=1      % required by arXiv

\documentclass[aps,prb,groupedaddress,showpacs,twocolumn]{revtex4}

\usepackage{graphicx}
\usepackage[latin1]{inputenc}
\usepackage{color}

\newcommand{\text}[1]{\ensuremath{{\rm{#1}}}}
\newcommand{\un}[1]{{\,\text{#1}}}

\begin{document}

\title{Characterisation of Ferromagnetic Contacts to Carbon Nanotubes} %
\author{D.~Preusche}
\author{S.~Schmidmeier}
\author{E.~Pallecchi}
\author{Ch.~Dietrich}
\author{A.~K.~H\"{u}ttel}
\author{J.~Zweck}
\author{Ch.~Strunk}
\affiliation{Institute for Experimental and Applied Physics,
University of Regensburg, 93040 Regensburg, Germany.}

\begin{abstract}
We present an investigation of different thin-film evaporated
ferromagnetic materials for their suitability as electrodes in individual
single-wall and multi-wall carbon nanotube-based spin devices. Various electrode
shapes made from permalloy ($\rm{Ni_{81}Fe_{19}}$), the diluted ferromagnet
$\rm{PdFe}$, and $\rm{PdFe/Fe}$ bilayers are studied for both their
micromagnetic properties and their contact formation to carbon nanotubes.
Suitable devices are tested in low-temperature electron transport measurements,
displaying the typical tunneling magnetoresistance of carbon nanotube pseudo
spin valves.
\end{abstract}


% 75.75.+a: Magnetic properties of nanostructures

% 75.70.Kw: Magnetic properties of thin films, surfaces, and interfaces
% 		Domain structure (including magnetic bubbles)

% 85.35.Kt: Nanotube devices

% 72.25.-b: Spin polarized transport (for ballistic magnetoresistance, see
%          75.47.Jn; for spin polarized transport devices, see 85.75.−d)

\maketitle

\section{Introduction}

Carbon nanotubes have been a widely investigated material for spintronics devices over the last years, which is above all due to their unique electrical properties, i.~e. high Fermi velocity, quasi one-dimensional ballistic transport, long spin lifetimes, and weak influence of nuclear spin.\cite{citeulike:1119796,NanospintronicsCNTferroKontosREVIEW} To obtain spin-dependent electron emission or absorption, a nanotube is typically contacted
by ferromagnetic electrodes. Thin-film ferromagnets diluted with Pd are
known to make a good electrical contact to individual carbon nanotubes,
accessing the Kondo\cite{KondoNiSWCNTLinedelof, mwcntkondoSchoenenberger} and
Fabry-Perot regimes.\cite{SWCNTMRfabryPerotMorpurgo} Pd is used for its contact
transparency to carbon nanotubes\cite{PdswcntballistictransparentContactDai} and
its transgression into a ferromagnetic phase with even small admixture of a
ferromagnet.\cite{PdFegiantparamagnetic} For both
multiwall\cite{MWCNTspinInjTranspKontos, mwcntkondoSchoenenberger,
NanospintronicsCNTferroKontosREVIEW, mwcntmagnetotransportMagneticEdge-domain}
and single wall carbon nanotube devices,\cite{SWCNTMRVgPdNiKontos,
SWCNTMRfabryPerotMorpurgo, KondoNiSWCNTLinedelof,
NanospintronicsCNTferroKontosREVIEW} the tunnel
magnetoresistance (TMR)\cite{Julliere1975225} has been observed. Any
industrial application of the
observed principles, however, will require a large degree of reproducibility
and control, in particular regarding the detailed micromagnetic
structure of the nanotube contact electrodes.

This work aims at an improvement of the switching properties of the
ferromagnetic
electrodes, i.\ e.\ obtaining reliable reversal of the magnetisation at
reproducible coercive field values, with a focus on PdFe alloys. Simultaneously,
we monitor the electrical
interface quality from ferromagnet to nanotube, including low-temperature
magnetotransport measurements. Here, reliable control of magnetic switching is
of particular
importance as there are no direct means to check the magnetic domain
structure once the sample is mounted in a cryostat for transport measurements.

\begin{figure}[ht]
\includegraphics[width=8.5cm,
angle=0,keepaspectratio]{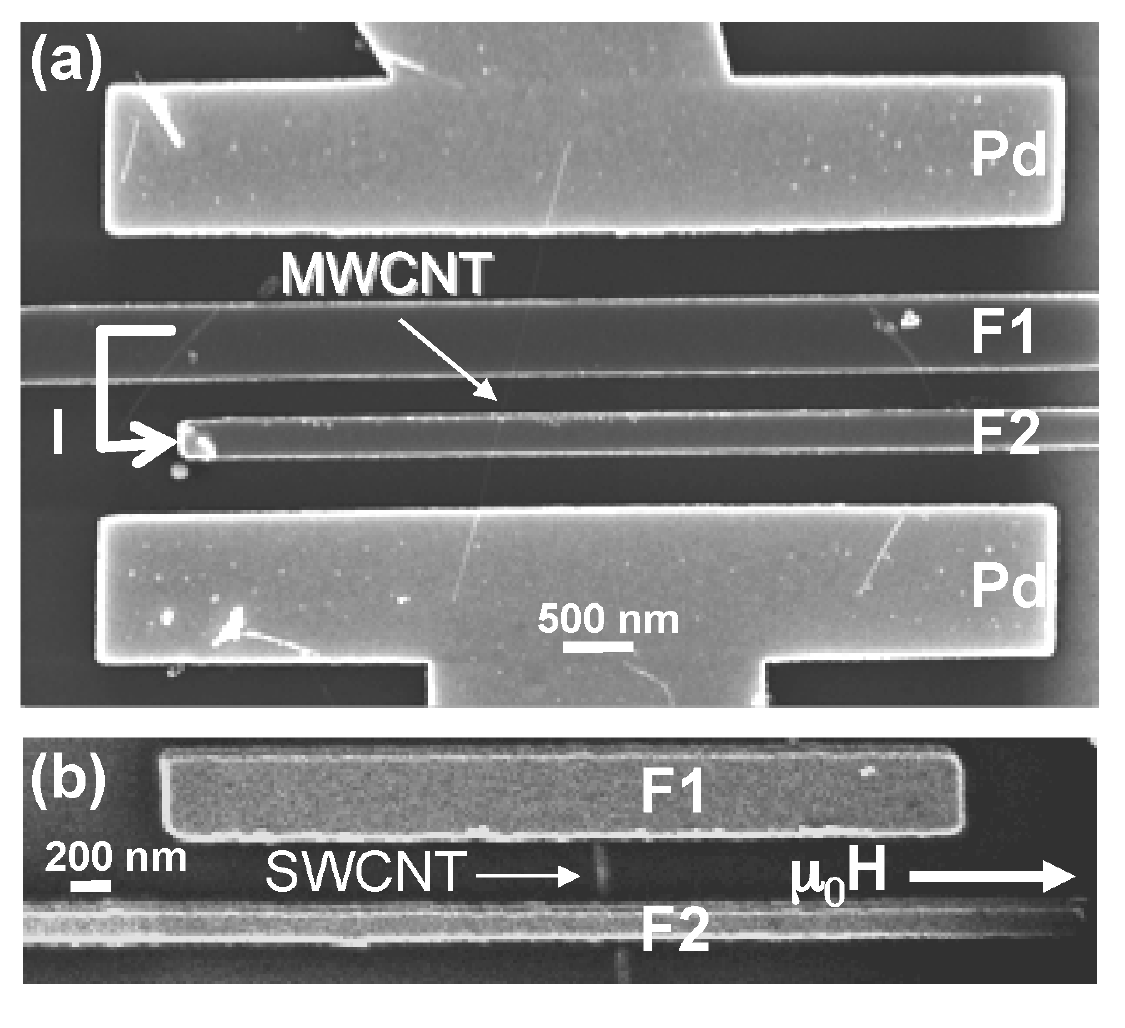}
\caption{
(a)
Scanning electron microscopy (SEM) image of a multiwall carbon nanotube (MWCNT)
contacted by two ferromagnetic electrodes, forming a pseudo spin
valve device. The electrodes F1 and F2 are fabricated from permalloy (Py, or
$\rm{Ni_{81}Fe_{19}}$); in addition two non-ferromagnetic Pd electrodes to
allow four-terminal measurements are shown.
(b)
SEM image of a single-wall carbon nanotube (SWCNT) contacted by two
ferromagnetic electrodes F1 and F2 made from the dilute ferromagnet
$\rm{Pd_{60}Fe_{40}}$. An arrow indicates the direction of an externally
applied magnetic field. Because of shape anitrosopy, the different electrode
aspect ratios result in a differing coercive field.
}\label{samplelayout}
\end{figure}
We investigate a typical pseudo spin valve
geometry,\cite{MWCNTspinInjTranspKontos, MRSWCNTCoKim,
mwcntmagnetotransportMagneticEdge-domain, SWCNTMRfabryPerotMorpurgo} as shown
in Fig. \ref{samplelayout}(a) for a multi-wall carbon nanotube (MWCNT) and in
(b) for a single-wall carbon nanotube (SWCNT).  Two ferromagnetic electrodes F1
and F2 are designed such that they have a different coercive field.  This is achieved here by making use of shape anisotropy, which can be tuned via the sample geometry.
In magnetoresistance measurements, a sufficiently large external magnetic field
is applied to saturate and align the magnetisation of both contacts in parallel
to it. A magnetic field sweep to opposite field direction will first switch the
magnetisation of the contact with smaller coercive field, thereby achieving antiparallel
configuration, and then the magnetisation of the second, resulting in
a parallel configuration with polarity opposite to the initial one.

\section{Lorentz microscopy imaging of the
magnetic switching of ferromagnetic electrodes}

For carbon nanotube magnetotransport experiments, the ferromagnetic contact
electrodes are required to have a difference in coercive field large enough to be
resolvable in transport measurements, e.g.\ $10\un{mT}$ or more. A second
requirement for reliable and reproducible magnetic switching is that the segment
of the electrode contacting the carbon nanotube be in a well-defined single
domain state to allow experimental control over injecting a spin-polarized
current. Transmission electron microscopy (TEM) in Lorentz microscopy
mode provides a powerful tool to investigate these properties of thin ferromagnetic films.

\subsection{Methods}

Lorentz microscopy\cite{LoMicroReview, LorentzMicroSpinValveChapman,
LorentzMicroPyCircSwitchingZweck} allows direct observation of the magnetic
domain structure of a ferromagnetic structure and its evolution in an external
magnetic field by sending a parallel (out of focus) electron beam through a magnetic specimen. The
deflection of the beam due to the Lorentz force can be visualized by defocusing
the objective lens.
At the walls enclosing a magnetic domain, the electrons are deflected as they transverse the ferromagnetic film and form either a convergent or a
divergent set of partial electron waves. Correspondingly, an increase or a
decrease of the electron beam intensity is detected at the
location of domain walls. It should be noted that not only domain walls but also other
variations in the magnetic induction within the specimen can be visualized by
this technique.

If the specimen is magnetized homogeneously, every transmitted electron will be deflected to the same side. In this case, a brightly contrasted feature forms on the edge of the specimen where the deflected beam through it overlaps with the undeflected beam passing next to it. On the opposite side, the deflection of the electrons will partially deplete the edge region which manifests in a dark contrast line on the detector.
\begin{figure}[ht]
\includegraphics[width=8.5cm, angle=0,
keepaspectratio]{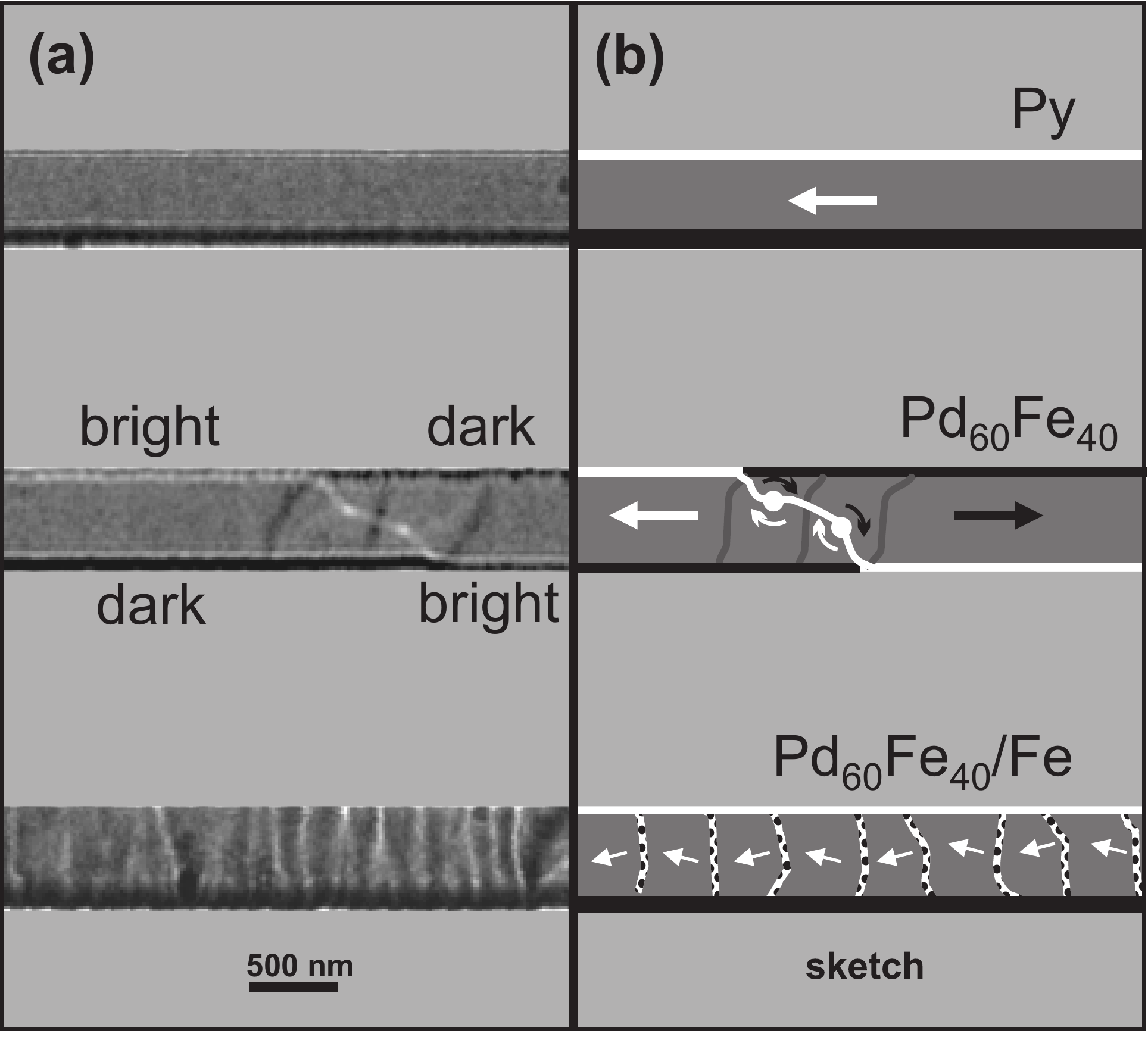}
\caption{
(a) Comparison of Lorentz microscopy images of rectangular  electrodes made from
different ferromagnetic materials. An external magnetic field is tuned close
to the coercive field. All strips have lateral dimensions of $500\un{nm}
\times 16\,\mu\text{m}$; the film thickness is $45\un{nm}$.
(b) Sketch pointing out the observed micromagnetic features. The
magnetisation direction (big black and white arrows) can be be determined by the
position of the dark and bright contrast. Solid black and white
lines, as seen in the $\rm{Pd_{60}Fe_{40}}$ strip are caused by magnetic domain
walls with full magnetisation reversal. Cross-tie domain walls (grey lines) enclose a magnetic vortex (white dot with curved white and black arrows). Weak contrast lines, drawn dotted in the
sketch, correspond to ripple domain walls with only small changes in
magnetisation direction (small white arrows).
}\label{FMstripsPdFeFePdVSPy}
\end{figure}
These dark and bright contrast features can, for instance, be seen within the permalloy strip shown in the topmost panel of Fig.~\ref{FMstripsPdFeFePdVSPy}. Magnetic domain walls, where spins oppose each
other in frustration, appear as dark or bright features within the strip area (see middle panel of Fig.~\ref{FMstripsPdFeFePdVSPy}). A detailed
discussion of Fig.~\ref{FMstripsPdFeFePdVSPy} will be given
below.

Driving the electron beam focus from above to below the sample plane reverses the Lorentz-force induced
contrast, providing a consistency test as to whether observed features are indeed of
magnetic origin. A magnetisation reversal due to an external magnetic field can be
detected by observing subsequent images during
a field sweep when a structure's edge contrast is inverted.
Note that on the borders of the observed ferromagnetic structures, the dark
Lorentz contrast appears more pronounced than the bright one. This is partly owed to a superposition with a
bright contrast all around the observed structures stemming from Fresnel edge diffraction. The magnetisation structure can therefore be read off most clearly
by tracing structures of dark edge contrast.

In the TEM used for the work at hand, an external magnetic
field $\mu_0 H$ can be applied only in parallel to the electron beam,
i.\ e.\ perpendicular to the sample plane. An in-plane field component $\mu_0 H_{||}=\mu_0 H\sin{\alpha}$ with respect to the sample plane can be tuned by maintaining this field at constant magnitude and tilting the sample by an angle $\alpha$.
At a maximum tilt angle of $\pm 25^\circ$, the in-plane
component reaches about half the value of the out-of-plane component.
Rotating the sample and thereby sweeping the in-plane field from saturation
through zero field to saturation in opposite direction allows to monitor the magnetisation reversal process of the ferromagnetic specimen.

Electron transmission microscopy requires samples to be prepared on a grid or thin film membrane with low electron beam contrast. The metal test structures were
patterned by electron beam lithography (EBL) and thermally evaporated in
vacuum onto $50\un{nm}$ thin low-stress PECVD silicon nitride membranes.
Owing to shape anisotropy, N\'{e}el walls, i.\ e.\ with the frustrated spins being confined to the film plane, occur in sufficiently thin
ferromagnetic films.
Above a material-specific thickness, which is smaller for weaker magnetisation,
Bloch walls dominate. Here, the spins form domain walls by gradually turning
out of the film plane. According to our experience, the threshold thickness of permalloy
(Py, $\rm{Ni_{81}Fe_{19}}$) films is about $50\un{nm}$. We therefore keep the film thicknesses below that value. Also the
large aspect ratio of film thickness compared to the lateral electrode
dimensions is expected to favour an in-plane orientation of the
magnetisation.\cite{MagneticDomains} We hence start our
discussion with the effect of this in-plane magnetic field component. In any
case, effects of magnetisation and magnetic field components
parallel to the electron beam are not imaged in Lorentz microscopy, due to the cross product
in the Lorentz force.

\subsection{Material dependence of the domain structure and magnetisation
reversal properties}

The investigated materials are chosen for their magnetic properties or expected
contact transparency to carbon nanotubes. We consider Permalloy
(Py, $\rm{Ni_{81}Fe_{19}}$) and the giant paramagnet Pd diluted with Fe, as we expect diluted ferromagnetic PdFe alloys to combine the benefits of the strong
ferromagnetism of iron and the transparent contacts of palladium -- carbon
nanotube interfaces.\cite{PdswcntballistictransparentContactDai,
NanospintronicsCNTferroKontosREVIEW} The composition of the PdFe
alloy is adjusted by setting the evaporation rates from two confocal thermal
evaporation sources appropriately. In addition, magnetic bilayer structures
with $10\un{nm}$ of $\rm{Pd_{60}Fe_{40}}$ and a
$35\un{nm}$-thick Fe layer -- to stabilize the magnetisation of the magnetically soft PdFe
by the magnetically hard Fe -- are discussed. To prevent
the iron from oxidation and for improved interface resistance with the Pd leads,
these bilayers were additionally capped with $5\un{nm}$ of Pd.

In Fig.~\ref{FMstripsPdFeFePdVSPy}(a), rectangular strips of equal dimensions
made from these three polycrystalline materials are compared. To allow comparison of the domain structure in different ferromagnets, the Lorentz microscopy images are
taken at the  field values close to magnetisation reversal in the specific material. We concentrate
initially on the
central segment of the strip, where the
contacted carbon nanotube would be placed in transport measurements. The majority
spin orientation and magnetic domain structure of this portion determines spin
orientation and polarisation of an injected current. The sketches of
Fig.~\ref{FMstripsPdFeFePdVSPy}(b) point out the relevant
magnetic features of the Lorentz images.

\subsubsection{Permalloy}

Permalloy (Py) was a chosen as a ferromagnetic material for its high magnetic
permeability, low coercive
field, and large magnetic anisotropy.\cite{MagneticDomains, Pydomainconfig,
PyStripSizeDepMagn} From the homogeneity of the entire Py strip
in Fig.~\ref{FMstripsPdFeFePdVSPy}(a) it can be concluded that the observed segment is in a
single domain state. In the Lorentz image the magnetic domain appears bordered by a continuous bright contrast line on the top side and a dark one on the bottom side.
The corresponding arrow in Fig.~\ref{FMstripsPdFeFePdVSPy}(b) symbolizes this
uniform magnetisation orientation of the single domain. Note that
the absolute direction of the magnetisation can only be determined from the dynamics during a full magnetic sweep by comparing subsequent Lorentz images.

\subsubsection{PdFe alloy}

In contrast to Py, the $\rm{Pd_{60}Fe_{40}}$ strip
segment in Fig.~\ref{FMstripsPdFeFePdVSPy}(a) is in a two-domain state at a small field increase beyond magnetisation reversal but not yet saturated. The black and white
arrows in the sketch of Fig.~\ref{FMstripsPdFeFePdVSPy}(b) highlight the
opposite
magnetisation orientation of the domains. The border contrast changes from bright to dark where the sharp white line connects to the strip boundary. In addition, three dark-contrasted, roughly parallel lines can be made out
crossing the bright line.
Such so-called cross-tie wall structures are typical for samples between N\'{e}el and Bloch
phase\cite{MagneticDomains} where a successive row of in-plane
(N\'{e}el) and out-of-plane (Bloch) type walls occurs. For our polycrystalline specimen, we expect a vanishing crystalline anisotropy,
so the magnetic dipolar(or shape) anisotropy is dominant. Due to the small magnetisation of $\rm{Pd_{60}Fe_{40}}$
compared to stronger ferromagnetic materials, also the shape
anisotropy is small. This means that even low magnetic fields
perpendicular to the film can lead to relatively large normal components $M_z$ of
magnetisation.
In any case, magnetic multi-domain and cross-tie wall
configurations of the $\rm{Pd_{60}Fe_{40}}$ strip render its magnetisation
configuration ill-defined for the operation of a CNT pseudo spin-valve. Furthermore, the specific domain wall pattern is often different for any two magnetic sweeps. Occurrence of magnetoresistance in a
pseudo spin valve fabricated from this material is therefore prone to a random domain
configuration and the position of magnetic pinning centers.

\subsubsection{PdFe alloy with a Fe magnetisation stabilisation layer}

The $\rm{Pd_{60}Fe_{40}/Fe}$ bilayer strips display a more uniform overall
magnetisation, as indicated by a continuous dark, respectively bright, contrast
along the entire borderlines of the strip. In addition, a fine structure appears in
Fig.~\ref{FMstripsPdFeFePdVSPy}(a), which signals the presence of ripple
domains, i.e. areas in which the spins deviate from the overall magnetisation
direction by a small angle as sketched in the bottom structure of
Fig.~\ref{FMstripsPdFeFePdVSPy}(b). The formation of ripple domains
is a well-known phenomenon for thin film materials with a high saturation
magnetisation like Fe,\cite{MagneticDomains} suggesting
that the Fe layer dominates the magnetic properties of the strip.

\begin{figure}[ht]
\includegraphics[width=7.2cm, angle=0,
keepaspectratio]{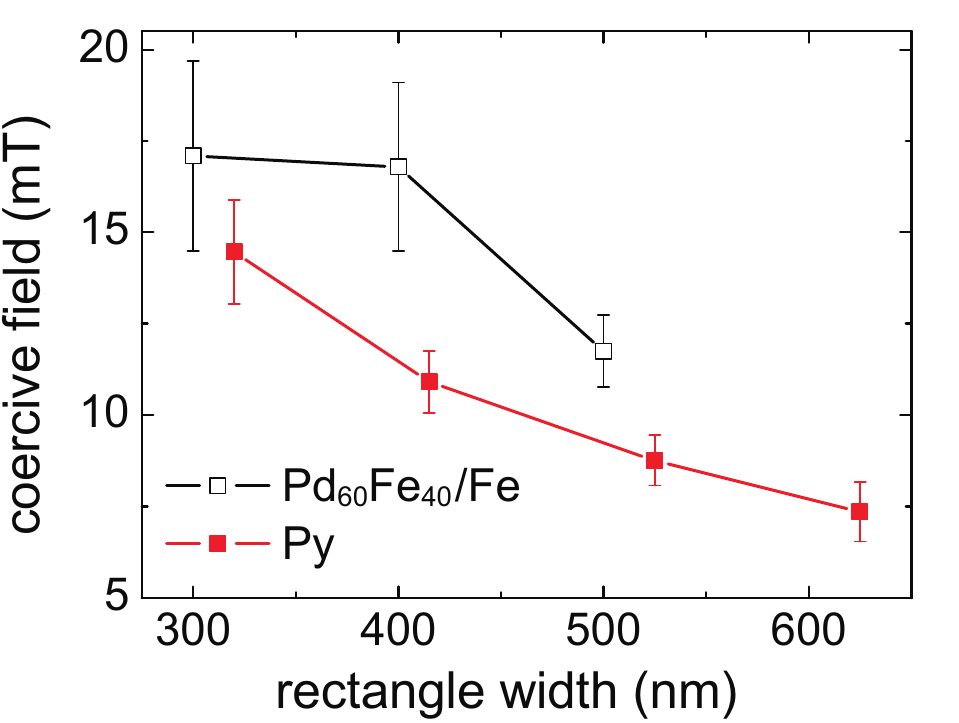}
\caption{
Comparison of the coercive field of Py and
$\rm{Pd_{60}Fe_{40}/Fe}$ strips.
Data points are retrieved from room temperature Lorentz microscopy measurements and averaged over three strips of same dimensions fabricated in the same EBL and
metallisation process.
While the absolute switching field is higher for strips of $\rm{Pd_{60}Fe_{40}/Fe}$
than of Py, the deviation from strip to strip is much larger. Py
displays a more reliable switching (smaller error bars).
}\label{MatComparePyVSPdFe-Fe}
\end{figure}
Fig.\,\ref{MatComparePyVSPdFe-Fe} compares the switching behaviour of
$\rm{Pd_{60}Fe_{40}/Fe}$ and $\rm{Py}$ strips of equal lateral dimension and total
thickness of $\rm{45~nm}$. While the $\rm{Pd_{60}Fe_{40}/Fe}$
magnetic bilayer switches at higher fields than permalloy, it does so with a
much higher margin of error.

\subsection{Shape dependence of the coercive field}

Next, we compare the magnetic switching behaviour of
rectangular, needle-shaped\cite{Femagnetic-switchingneedlerectangle,
LoMicroPyNeedlesRectangles} and spoon-shaped\cite{michaelhuber} electrode
structures in order to identify the optimal geometry parameters for magnetic
switching reproducibility and controllability. It is well-known that by choice
of very high aspect ratio of length to width, the smaller of the two parameters
dominates the switching behaviour via shape
anisotropy. By setting the length to $16\,\mu\text{m}$ for all structures,
investigations can be reduced to varying the electrode widths in the
range from $200\un{nm}$ up to $\sim 1\,\mu\text{m}$. Lorentz microscopy allows to determine whether a structures is of suitable dimensions to be in a single domain configuration.

\subsubsection{Rectangles}

\begin{figure}[ht]
\includegraphics[width=8.5cm, angle=0,
keepaspectratio]{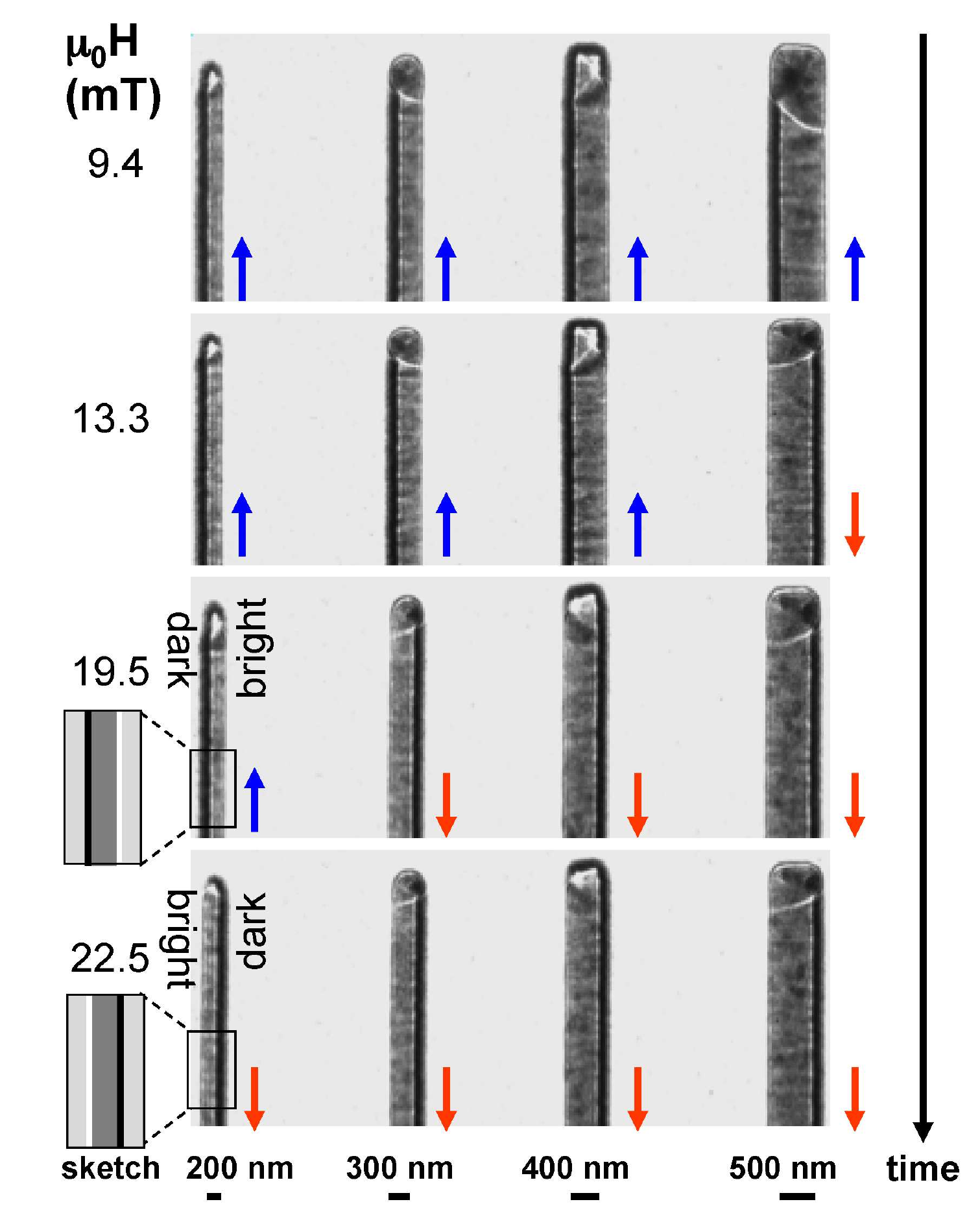}
\caption{
Lorentz microscopy images of four $\rm{Pd_{60}Fe_{40}/Fe}$ ($\rm{10~nm/35~nm}$
thick) strips of different width ($200\un{nm}$, $300\un{nm}$,
$400\un{nm}$, $500\un{nm}$). The length ($16\,\mu\text{m}$)
is chosen large enough for the width only to govern the shape anisotropy.
The strips have been fully saturated in a  magnetic field parallel to the long
strip axis (blue arrows, top row). When applying an increasing external
field in opposite direction (subsequent lower image rows), magnetic switching
takes place, see the red and blue arrows next to the strips.
The required coercive field clearly decreases with increasing strip width. The
black arrow indicates how images were taken subsequently in time during a
single magnetic field sweep. Insets: sketches of the observed edge features, cf.
Fig.~\ref{FMstripsPdFeFePdVSPy}.
}\label{FMstripsswitching}
\end{figure}

Fig.~\ref{FMstripsswitching} shows four $\rm{Pd_{60}Fe_{40}/Fe}$
contact strips of different width at different values of the external magnetic
field $\mu_0 H_{||}$. Coming from saturation, where all strip magnetisations are
aligned along $-\mu_0 H_{||}$ (blue arrows, top image row), the external field
is swept to opposite field direction $\mu_0 H_{||}$. As indicated by the arrow
on the edge, subsequent image rows are recorded at increasing field values during a magnetic field sweep. The
images show that the rectangular strips flip their magnetisation consecutively
at coercive fields increasing in order of decreasing strip width.

An interesting feature visible in Fig.~\ref{FMstripsswitching} is the formation
of a magnetic end domain at the upper end of the electrode. Its evolution can
be monitored by following the magnetic features at increasing field
values. It indicates that the magnetic end domains nucleate the magnetic
switching of the whole strip. Otherwise, the strips all show a clear single
magnetisation direction as desired for a pseudo spin valve contact electrode.

\subsubsection{Needle-shaped structures}

\begin{figure}[ht]
\includegraphics[width=8.5cm, angle=0,
keepaspectratio]{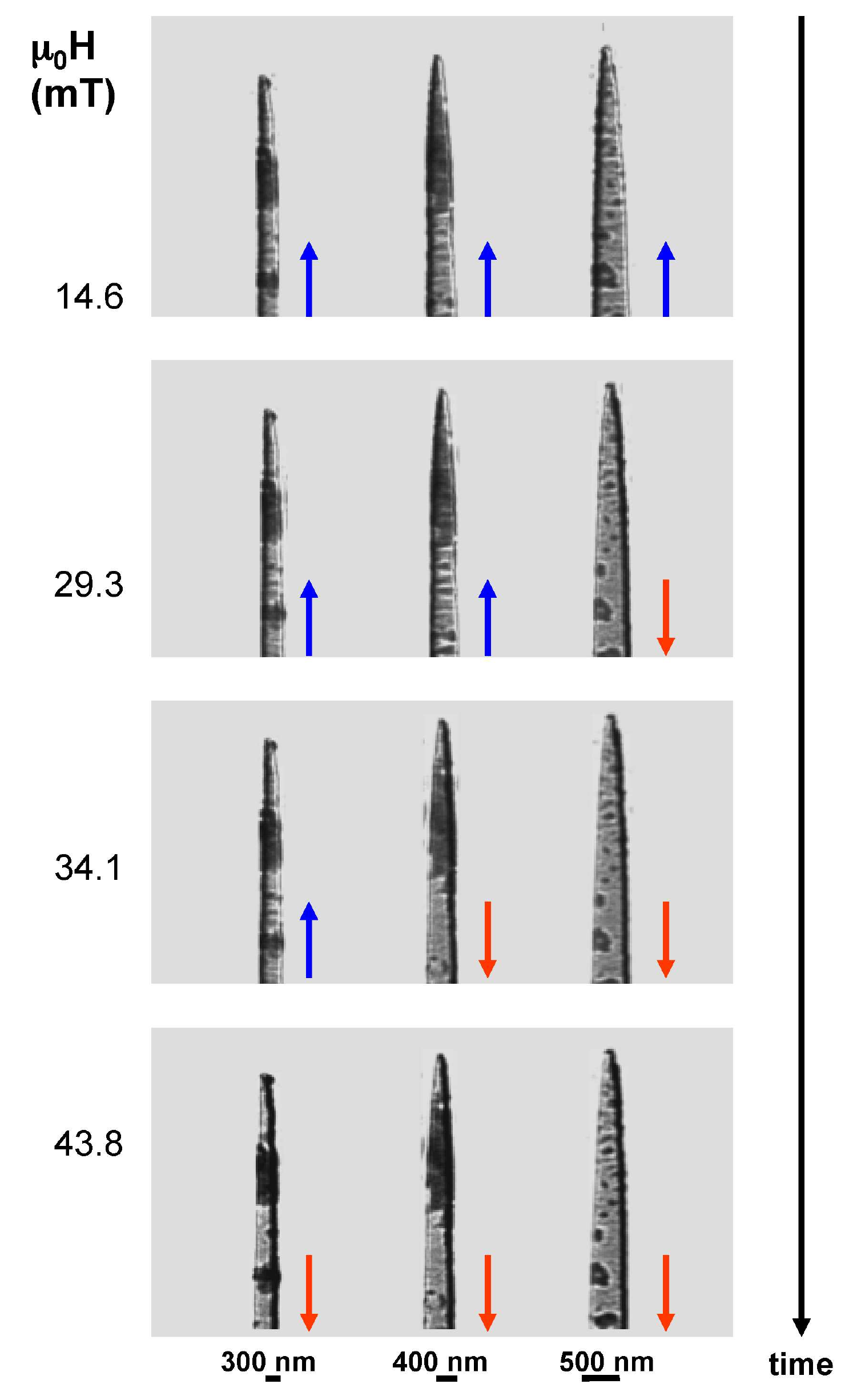}
\caption{
Lorentz microscopy images detailing the magnetisation reversal of needle-shaped
 $\rm{Pd_{60}Fe_{40}/Fe}$ ($\rm{10~nm/35~nm}$) structures differing in width
($300\un{nm}$, $400\un{nm}$, $500\un{nm}$), analogous to the data of
Fig.~\ref{FMstripsswitching}. From upper to lower image row, an increasing
magnetic field directed opposite to the original saturation magnetisation is
applied. The irregular spots on the image stem from non-magnetic process contamination.
}\label{FMneedleswitching}
\end{figure}
In Fig. \ref{FMneedleswitching}, three $\rm{Pd_{60}Fe_{40}/Fe}$ needle-shaped
structures of different width are Lorentz-imaged. A magnetisation reversal
analogous to Fig.~\ref{FMstripsswitching} is found. Compared to rectangles, switching in needles is also governed by the structure width but occurs at higher coercive field. This may be explained by fewer magnetically frustrated spins within the pointed structure tips, i.\ e.\ fewer and smaller end domains, see Fig.~\ref{FMneedleswitching} compared to the rectangles of
Fig.~\ref{FMstripsswitching}. As the magnetisation reversal is initiated by
these end domains, their suppression also should translate into higher magnetisation stability.

\subsubsection{Spoon-shaped structures}

\begin{figure}[ht]
\includegraphics[width=8.5cm, angle=0,
keepaspectratio]{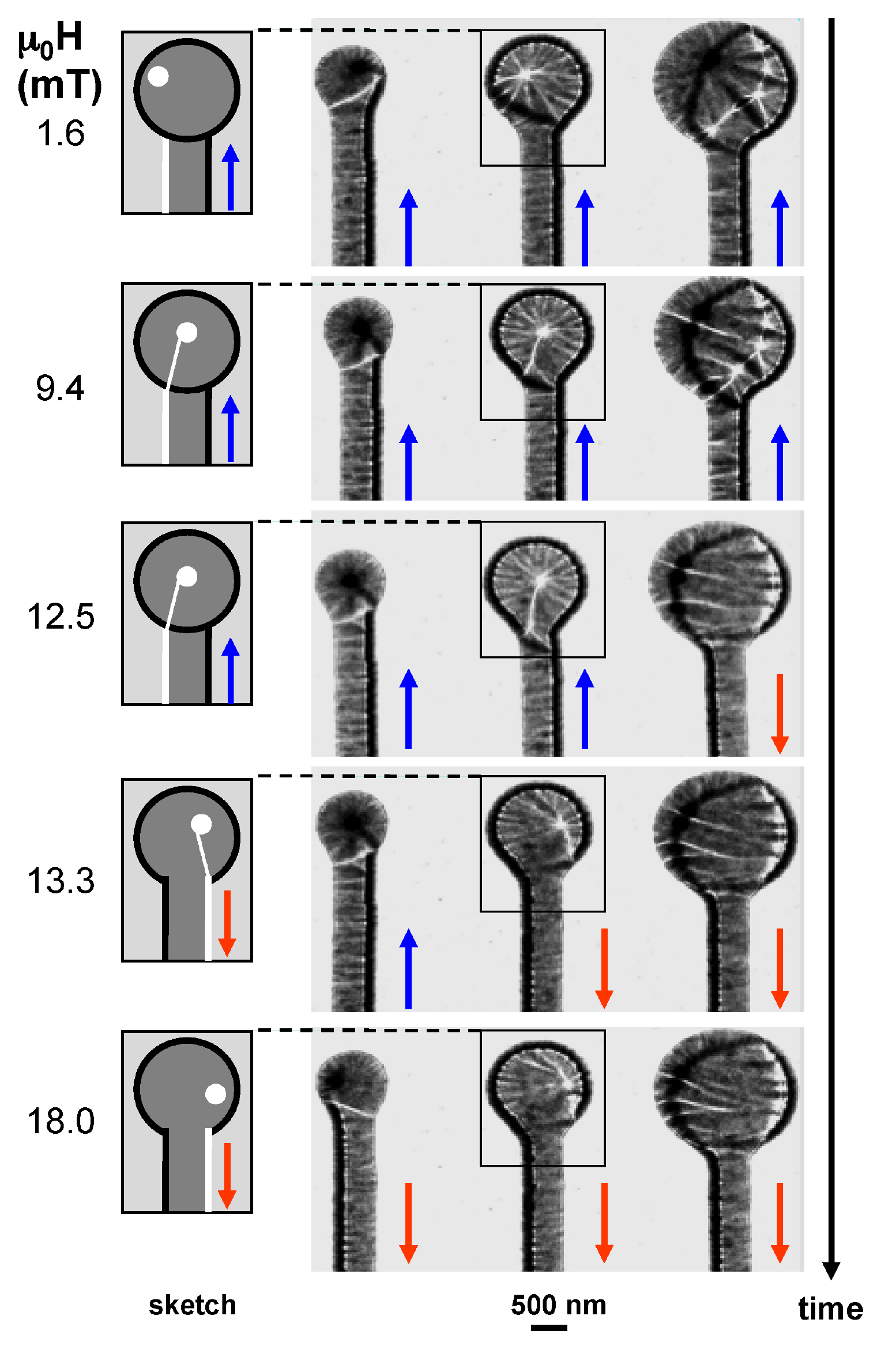}
\caption{
Lorentz microscopy images showing the magnetisation reversal of
three spoon-shaped structures. Each of the identical, $\rm{16}$ $\rm{\mu m}$
$\rm{\times}$ $\rm{500}$ $\rm{nm}$, $\rm{Pd_{60}Fe_{40}/Fe}$ rectangular strips
has a disc-shaped structure of different diameter ($\rm{1.0}$ $\rm{\mu m}$,
$1.5\,\mu\text{m}$, $\rm{2.0}$ $\rm{\mu m}$) attached at its upper end. The
lower part of the strip (not shown, at large distance) has a rectangular ending.
Inside the disk, the contrasts structure indicates the formation of a magnetic
vortex. The magnetic switching of the strip with increasing external
field (descending in image rows) is triggered when a domain wall originating from this vortex is driven into the rectangle.
}\label{spoons}
\end{figure}
In Fig.~\ref{spoons}, TEM Lorentz microscopy images of spoon-shaped structures, i.~e. rectangular strips with attached disk,
are presented. The disk diameter dependence of the magnetisation
reversal of attached rectangular strips is investigated. All
strips have equal dimension, only the disk diameters are varied.

In a disk made from a ferromagnetic material, the spins can align in a vortex
configuration.\cite{MagneticDomains} Depending on whether the vortex chirality
is clockwise or
counter-clockwise, the vortex core appears in the Lorentz image as either a dark spot (structure in the
first column of Fig.~\ref{spoons}) or as a bright spot (second column).
At zero external magnetic field, the vortex core is near the center of the
disk. An external magnetic field along the strip ``shoves'' the bright
(dark) vortex core to the right-hand side (left-hand side).

This behaviour can be observed most clearly in the example of the second column
of Fig.~\ref{spoons}, where the magnetisation reversal of
a ``spoon'' with a $1.5\,\mu\text{m}$ diameter disc is shown. A domain wall, seen
as a bright line in the images, originates from the vortex core for all
magnetic field values. Sweeping up from negative values of $\mu_0 H$, the vortex core is still shifted at  $+1.6\un{mT}$ to the
left-hand side, owing to the magnetic remanence of the structure. The domain
wall ends at the left-hand side of the disk. A field of $9.4\un{mT}$ is
required to counter the magnetic remanence so the vortex is moved to the
disk center. The white domain wall ends now near the joining of disk and
strip as the domain to its left is expanding under the effect of the increasing magnetic
field. At $12.5\un{mT}$ the vortex core is shifted to the right-hand side and
the domain left to the said white domain wall is on the brink of expanding into
the rectangular strip beneath; a small increase to $13.3\un{mT}$ is
sufficient to cause the
domain wall to pass through the rectangular strip, thereby initiating a magnetisation
reversal. At $18.0\un{mT}$, the vortex core is almost 'pushed' out of the
disk and the entire spoon-shaped structure is saturated.
The switching field of the total spoon structure is lowered when the disk diameter is increased. We conclude that the disk can thus trigger the magnetisation
switching of the strip structure in a controlled way, independent of random end
domains. However, as can be seen in the third column of Fig.~\ref{spoons}, at
large disk diameters no clear vortex forms but strong ripple domain structures
typical for Fe thin films appear,\cite{FeLorentzmicroscopyRippledomain}
counteracting the expected stabilising effect on magnetic switching.

\begin{figure}[ht]
\includegraphics[width=7.7cm, angle=0,
keepaspectratio]{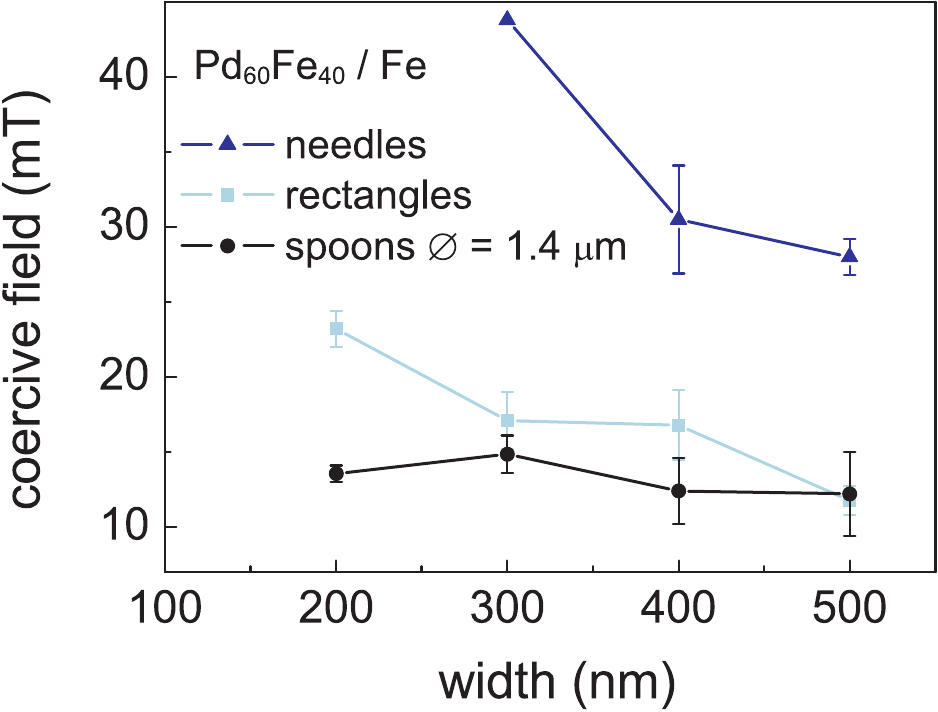}
\caption{
Shape influence on coercive field. The coercive field obtained from Lorentz
microscopy images is plotted for spoon-shaped, rectangular, and needle-shaped
structures made from $\rm{Pd_{60}Fe_{40}/Fe}$. In the case of the spoon shapes,
the $\rm{16}$ $\rm{\mu m}$-long rectangular 'handles' are topped with identical
disks of diameter $\rm{1.4}$ $\rm{\mu m}$. For all structures, the (handle)
strip width is varied (x-axis in the graph). Coercive field values are averaged
over 4 magnetic field sweeps of the same structure and shown with respective
error bars. The coercive field of the needles shows the strongest increase with
decreasing width. Whereas the rectangles still display a sizeable strip width
dependence of the coercive field, the disk dominates the magnetic switching of
the spoon structures, nearly equalizing the coercive field for all strip widths.
}\label{spoonscomparison}
\end{figure}

Both effects are illustrated by the data of Fig.~\ref{spoonscomparison}, where
we plot the coercive fields of strips ranging from $200\un{nm}$ to
$500\un{nm}$ with and without an attached disc of diameter $1.4\,\mu\text{m}$, or a needle tip.
The disc clearly dominates the switching of the spoon-shaped structures and equals the
coercive field of all strips around $12.5\un{mT}$. However, for larger structures a stronger scatter in measured values can be observed. The needles show the
strongest width dependence on the coercive field.

\subsubsection{Effect of ferromagnetic supply lines}

\begin{figure}[ht]
\includegraphics[width=8.5cm, angle=0,
keepaspectratio]{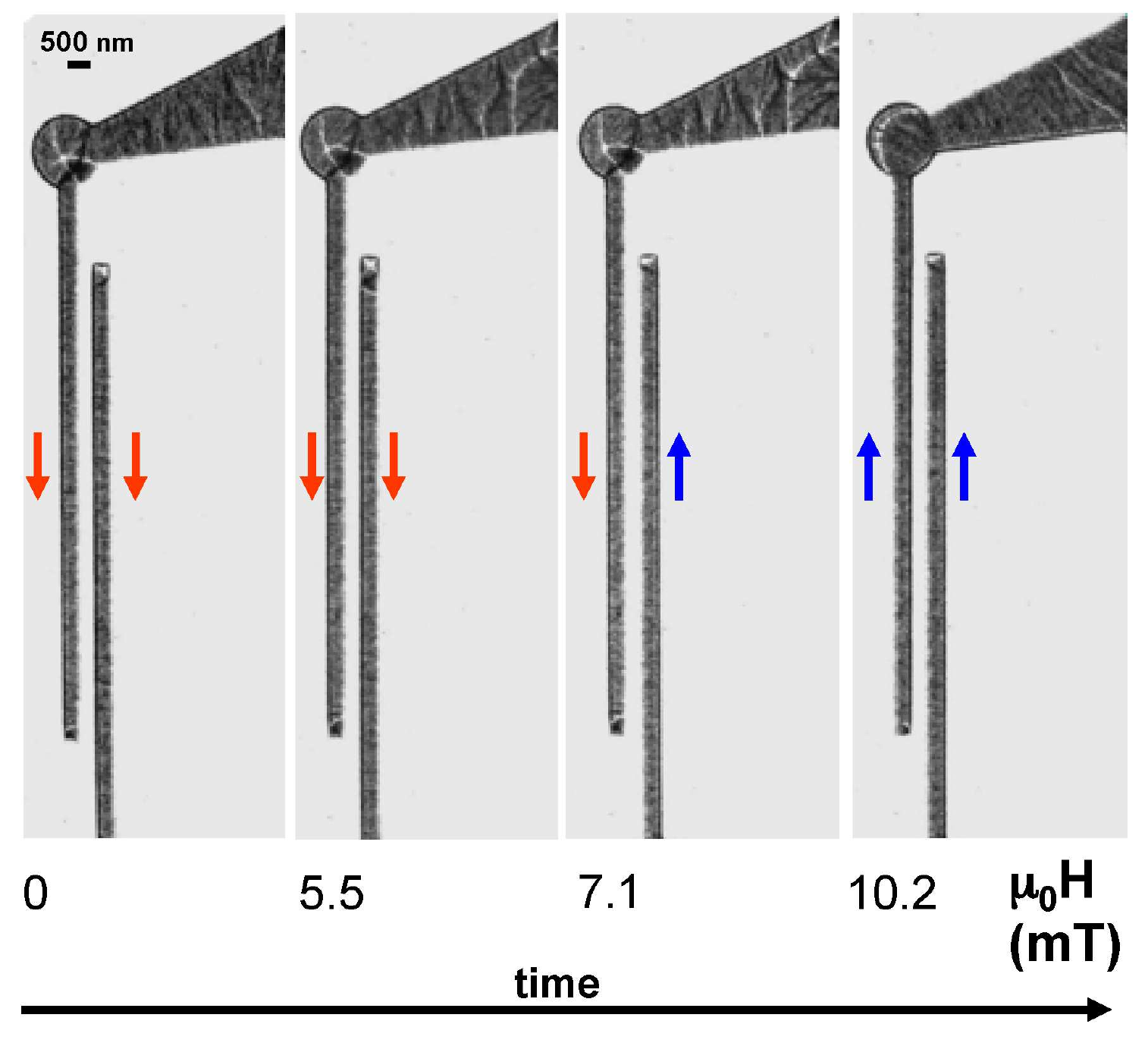}
\caption{
Lorentz microscopy images of a detached rectangular electrode compared to an
electrode with attched large-area ferromagnetic leads. The electrode width is in both
cases $500\un{nm}$. Following the images from left to right, it can be observed how an external magnetic field initiates a magnetisation reversal. The
ferromagnetic leads change the switching field since
the magnetisation reversal can be initiated by the increasing external field
driving a domain wall from the lead into the rectangular electrode.
}\label{FMstripsswitchingContact}
\end{figure}

Fabricating contact electrodes together with supply lines
and bond pads in a single metallisation process would simplify chip processing significantly. To investigate the effect of attaching large ferromagnetic supply
lines on the magnetisation reversal of the contact electrodes, two
identical $\rm{Pd_{60}Fe_{40}/Fe}$ strips with and without attached typical supply
lines and bond pad (not shown) geometry attached are compared in Fig.~\ref{FMstripsswitchingContact}. To reduce the magnetic coupling between electrode and supply lines a disk is placed at their junction. This disk typically forms a strongly diameter-dependent vortex, as discussed above.

Coming from saturation at high negative field, at $0\un{mT}$ (left panel in
Fig. \ref{FMstripsswitchingContact}) both contacts have their magnetisation
aligned downwards (red arrows). At $7.1\un{mT}$, initiated by edge domains, the
strip without attachment aligns along the external field (blue arrow). The strip with attachment follows at a coercive field of $10.2\un{mT}$ -- in this particular case at a higher rather than at lower field value, as also
observed with an attached disc structure. Still, there
is a difference in coercive field of $3.1\un{mT}$  indicating that the magnetisation reversal is triggered by the  multi-domain configuration in the supply
lines. Switching is rendered irreproducible due to the random domain structure in the bulky supply line.

\subsection{Temperature dependence}

\begin{figure}[ht]
\includegraphics[width=7.2cm, angle=0,
keepaspectratio]{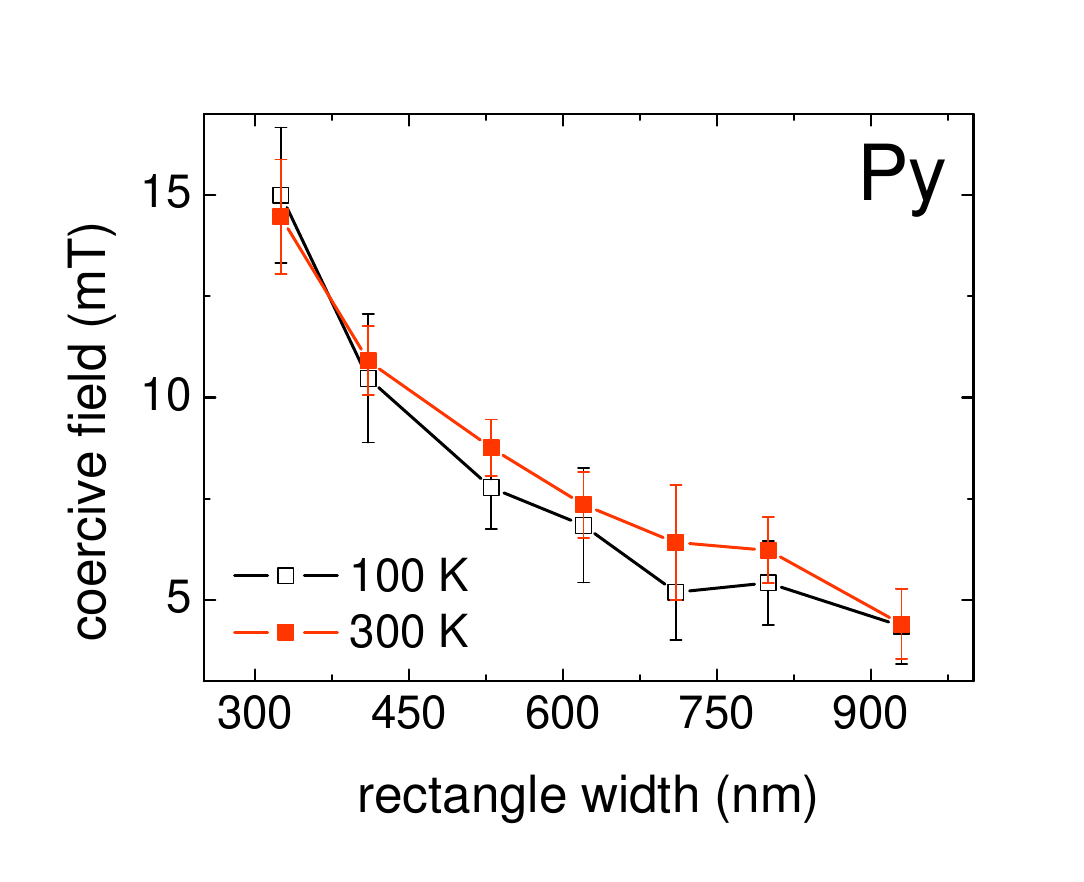}
\caption{
Comparison of the coercive field of rectangular Py strips of different width
and with a length of $16\,\mu\text{m}$ at room temperature ($300\un{K}$, red) and
close to liquid nitrogen temperature ($100\un{K}$, white). Data points are
retrieved from Lorentz microscopy measurements, averaging over three rectangular
Py strips of same dimension fabricated in the same EBL and metallisation
process.
}\label{FMstripsswitchingTemperature}
\end{figure}

Low temperatures are still an important prerequisite to many fundamental
studies of spintronics devices. This raises the question whether the results
obtained by room temperature Lorentz microscopy remain valid in the low-temperature regime. For this purpose,
Fig.~\ref{FMstripsswitchingTemperature}
compares the coercive fields of rectangular Py strips obtained at room
temperature with those at $100\un{K}$, the lowest accessible temperature in our
TEM sample holder. Coercive field values agree, within the margin of error, for both temperatures, justifying an extrapolation of our findings to the low-temperature limit.

\section{Application in magnetotransport measurements}

For electronic transport experiments, we have fabricated SWCNT samples with
rectangular electrodes made from $\rm{Pd_{60}Fe_{40}}$ as well as MWCNT samples
with electrodes made from $\rm{Pd_{80}Fe_{20}}$, Py and a
$\rm{Pd_{60}Fe_{40}/Fe}$ magnetic bilayer.  The magnetoresistance was measured
both in an in-plane magnetic field along the electrodes and also in a perpendicular, out-of-plane field.

\subsection{Fabrication of nanotube devices}

SWCNTs have been grown by
chemical vapour deposition (CVD) on highly p-doped Si wafers with $300\un{nm}$ thermal oxide. The substrate can serve as backgate electrode. As feed gas we use a mixture of methane
and hydrogen at $900^\circ\un{C}$ with $\rm{Fe(NO_3)_3\cdot9H_2O}$, $\rm{Al_2O_3}$
and $\rm{MoO_2(acac)_2}$ as catalyst.\cite{growth} Best
results were obtained with a $\rm{CH_4}$ flux of $1.5\un{sccm}$ and
a $\rm{H_2}$ flux of $0.23\un{sccm}$ although these parameters seem
to be specific for the individual setup.

For the purposes of fabricating an individual SWCNT device, defect-free, clean
nanotubes were located with respect to pre-patterned alignment markers using either scanning electron microscopy (SEM) or atomic force
microscopy (AFM). In addition, high purity arc-discharge grown MWCNTs with
typical diameters around $\rm{10~nm}$ were dispersed on wafers from solution and localized similarly. The fabrication of contact electrodes was adapted to
the nanotube location by electron beam lithography and thin film deposition.

\subsection{Room-temperature interface resistance of different ferromagnetic
electrode materials to carbon nanotubes}

While the magnetic domain structure of the contact electrodes is one defining
factor for the performance of a CNT pseudo spin valve device, the
nanotube contact resistance is another important material selection criterion.

A severe disadvantage of Py electrodes was their lack in interface quality to
carbon nanotubes. In most devices, contact resistances of several
 $\rm{M\Omega}$ were observed. Annealing for about $60\un{s}$ at
$500^\circ\un{C}$ and $10^{-4}\un{mbar}$ improved the room temperature
resistance of MWCNT samples temporarily; applying, to the same end, a large bias
voltage of $2-10\un{V}$ led to a significant drop of the resistance of the
tube-metal interface. However, the interface deteriorated again
on a scale of several days. Even while temporarily decreased the sample resistance was marked by strong fluctuations and abrupt jumps.

Palladium is known to form high-transparency contacts to carbon
nanotubes.\cite{FETcntPdDai} Correspondingly it has also been shown that
palladium-diluted ferromagnetic materials inherit this
property.\cite{SWCNTMRfabryPerotMorpurgo, SWCNTMRVgPdNiKontos, MRSWCNTCoKim,
KondoNiSWCNTLinedelof} Yet we found that even four electrodes
patterned on the same tube in the same lithography and $\rm{Pd_{60}Fe_{40}}$
evaporation process can have highly different room temperature contact
resistances, ranging from $\rm{20~k\Omega}$ to the $\rm{\rm{M\Omega}}$
range. While some CNT devices with PdFe contacts yielded transparent contacts to
carbon nanotubes, e.g. accessing the Kondo regime, the controllability of the magnetic domain configuration in this material remains insufficient.

In order to preserve the benefit of the low-resistive PdFe-CNT interface and
to simultaneously obtain strong in-plane magnetisation, the
aforementioned magnetic bilayer films consisting of $10\un{nm}$
$\rm{Pd_{60}Fe_{40}}$ and a $\rm{35~nm}$-thick Fe layer were used.
The contacting of carbon nanotubes with $\rm{Pd_{60}Fe_{40}/Fe}$ electrodes
proved to be less reliable than with $\rm{Pd_{60}Fe_{40}}$ only. Resistances were typically in the $\rm{M\Omega}$-range. With an annealing step, the
resistance of some samples could be reduced to room temperature values
of down to $120\,\text{k}\Omega$. The noise level was high compared to the
other two investigated contact materials. Characteristic for these samples
was the occurrence of random telegraph noise-like resistance jumps,
exceeding any magnetoresistance features by a factor of typically 2-3.

\subsection{Low-temperature
magnetoresistance measurements on SWCNTs}

\subsubsection{TMR in a parallel magnetic field}

\begin{figure}[ht]
\includegraphics[width=8.5cm, angle=0,
keepaspectratio]{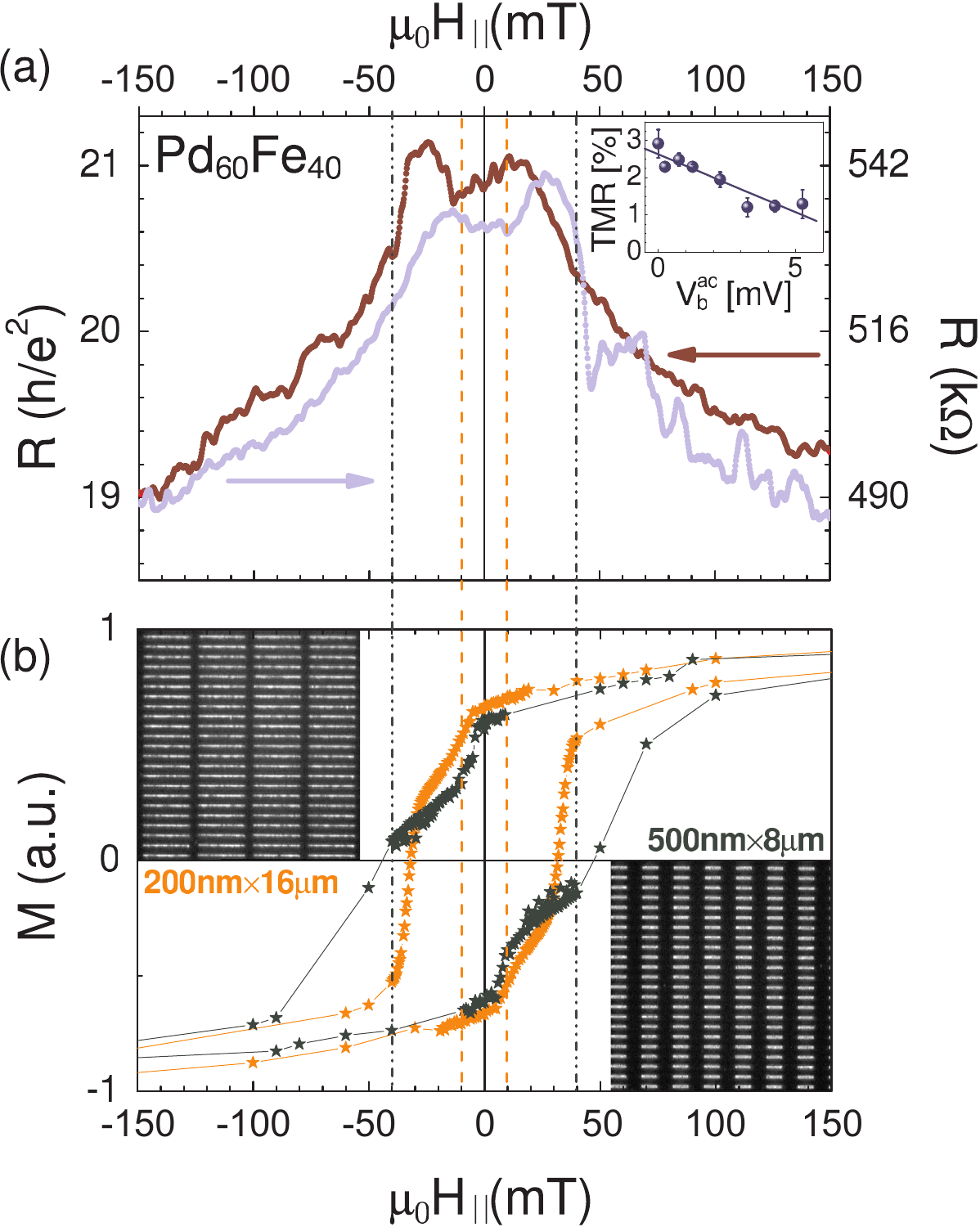}
\caption{%
(a)
Differential 2-terminal resistance $R=dV/dI$ of a SWCNT pseudo spin valve device
measured with lock-in techniques at $\rm{4.2~K}$ as a function of an in-plane magnetic field
$\mu_0 H_{||}$
along the electrodes long axis. The ac bias was $\rm{3~mV}$.
A TMR feature of $1.8\%$ appears in upsweep (light blue curve) and
downsweep (dark red curve) in the intervals of $\pm[10\un{mT},40\un{mT}]$.
The wide ($500\un{nm}\times 8\,\mu\text{m}$, marked in dark green)  strip switches at smaller field
than the narrow ($200\un{nm}\times 16\,\mu\text{m}$, marked in orange)
$\rm{Pd_{60}Fe_{40}}$ strip. Dashed orange and dark green
lines mark the magnetisation switching of the narrow and wide electrodes. Inset: TMR as
function of applied ac bias.
(b)
Magnetisation hysteresis $M(\mu_0 H_{||})$ of large arrays of
$\rm{Pd_{60}Fe_{40}}$ strips, with identical dimensions
as the contacts in (a), detected with a SQUID at $T=2\un{K}$. The saturation magnetisation is normalised to 1 and corresponds to $\rm{1.5\times 10^{-4}~emu}$ for the narrow strip array (orange) and $\rm{0.75\times 10^{-4}~emu}$ for the wide strip array (dark green).
Insets: SEM images of the samples.
}\label{TMRSWCNTparaSQUID}
\end{figure}
Fig.~\ref{TMRSWCNTparaSQUID}(a) shows the two-terminal resistance of a SWCNT
pseudo spin valve with $\rm{Pd_{60}Fe_{40}}$ rectangular electrodes. It is
plotted  as a function of the external magnetic field during a field upsweep
(light blue curve) and downsweep (dark red curve) from saturation of the electrodes to
saturation in opposite field direction. The external magnetic field was aligned
along the length of the electrode strips. On top of background features, a
positive TMR signal emerges as a step-like increase in resistance within an interval of $\rm{\pm[10~mT,~40~mT] }$. It arises from the relative magnetisation orientation of narrow  and wide  electrodes.

The amplitude of the tunneling magnetoresistance $\text{TMR} =
\left(R_{\uparrow\downarrow}-R_{\uparrow\uparrow}\right)/R_{\uparrow\uparrow}$
was inferred from measurements of the two-terminal resistances
$\rm{R_{\uparrow\downarrow}}$ and $\rm{R_{\uparrow\uparrow}}$ in parallel and
antiparallel magnetisation orientation of the two ferromagnetic electrodes.
We obtain a TMR effect of up to $3\%$ on top of a larger background magnetoresistance that is probably due to electron-magnon scattering. In some cases we observed a less significant maximum before reaching zero field, see e.g. the downsweep in Fig.~\ref{TMRSWCNTparaSQUID}(a). Its origin remains unclear at present. However, it seems to be absent when applying an out-of-plane field, compare Fig.~\ref{TMRSWCNT}.

The inset of Fig.~\ref{TMRSWCNTparaSQUID}(a) shows that the TMR drops with
increasing the applied dc bias. This is to be expected because, in addition to
the increased current, the differential conductance $dI/dV\equiv
I_\text{ac}/V_\text{ac}$ is effectively averaged over a larger source-drain
voltage $V_\text{sd}$ window.\cite{SWCNTMRfabryPerotMorpurgo}

\subsubsection{magnetisation reversal from SQUID measurements of
ferromagnetic strips}

To confirm that the observed magnetoresistance features indeed originate from
magnetic switching of the electrodes, we determined the coercive fields of typical thin-film contact electrodes at cryogenic temperature as a control experiment.
Fig.~\ref{TMRSWCNTparaSQUID}(b) shows
the hysteresis in magnetisation $M(\mu_0 H_{||})$ of narrow (orange) and wide
(dark green) rectangles during up- and downsweep of an external magnetic field
parallel to the strip axis. The signal was detected using a
commercial SQUID magnetometer. To obtain a sizable
magnetic moment, arrays of several ten thousand
identical strips were EBL-patterned as shown in the insets of Fig.
\ref{TMRSWCNTparaSQUID}(b).

The curves clearly show more features than the sharp and rectangular hysteresis
which would be expected from an ideal single domain switching event along a magnetically easy axis. The multiple features
suggest multiple domain switching, but may also stem partially from
end domain configurations, which do not impact the spin injection and
detection in the tube (see e.g. the Lorentz microscopy images of
Fig.~\ref{FMstripsswitching}). Still, when comparing the SQUID data
with the magnetotransport measurements, as indicated by the dashed dark green and
orange lines in Fig.~\ref{TMRSWCNTparaSQUID}, the switching of magnetoresistance
for both sweep directions can be matched to the sharp features in the
magnetisation curves. This is consistent with our findings from Lorentz
microscopy that the wide electrode (dark green) has lower coercive field than the narrow one (orange) by shape anisotropy.

\subsubsection{TMR in a perpendicular magnetic field}

\begin{figure}[ht]
\includegraphics[width=8.5cm,
angle=0,keepaspectratio]{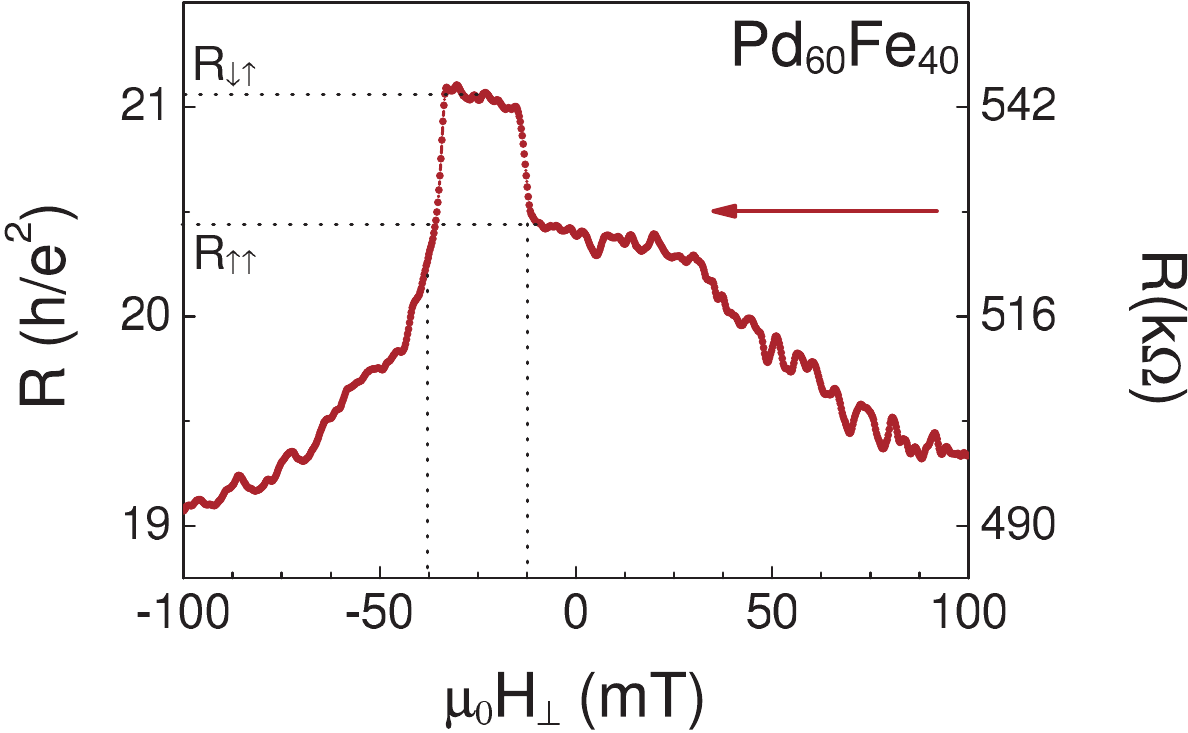}
\caption{
Differential 2-terminal resistance $R=dV/dI$ of the same SWCNT device as in Fig.
\ref{TMRSWCNTparaSQUID}  at $4.2\un{K}$, this time applying an out-of-plane
magnetic field $B_\perp$ and an ac bias of $3\un{mV}$. TMR features of $3\%$
appear in downsweep of the magnetic field  within the interval of
$[10\un{mT}, 40\un{mT}]$.
}\label{TMRSWCNT}
\end{figure}

Fig.~\ref{TMRSWCNT} displays an exemplary
magnetoresistance downsweep on the same sample as shown in Fig.
\ref{TMRSWCNTparaSQUID}, in a different cooling cycle, now applying a magnetic
field perpendicular to the sample plane. The resulting magnetoresistance curve
displays a lower noise level and a clearer
TMR feature with steeper flanks of up to $3\%$ is observed. This finding stands in contradiction to the assumption that the out-of-plane magnetisation
component is negligible in thin ferromagnetic films. It is consistent, however, with
our observation of cross-tie domain walls in these films in Lorentz microscopy,
hinting at a non-negligible out-of-plane magnetisation.
The switching is sharper than in parallel field, suggesting that the
strips flip from one single-domain state to the other with fewer or no
intermediate multi- or ripple domain states.
The easy out-of-plane magnetic axis may also be the cause for the higher
noise-level in parallel field magnetoresistance sweeps, due to a less
well-defined and potentially fluctuating domain configuration.

\subsection{Low-temperature magnetoresistance measurements  on MWCNTs}

\subsubsection{TMR in a parallel magnetic field}

\begin{figure}[ht]
\includegraphics[width=8.5cm, angle=0,
keepaspectratio]{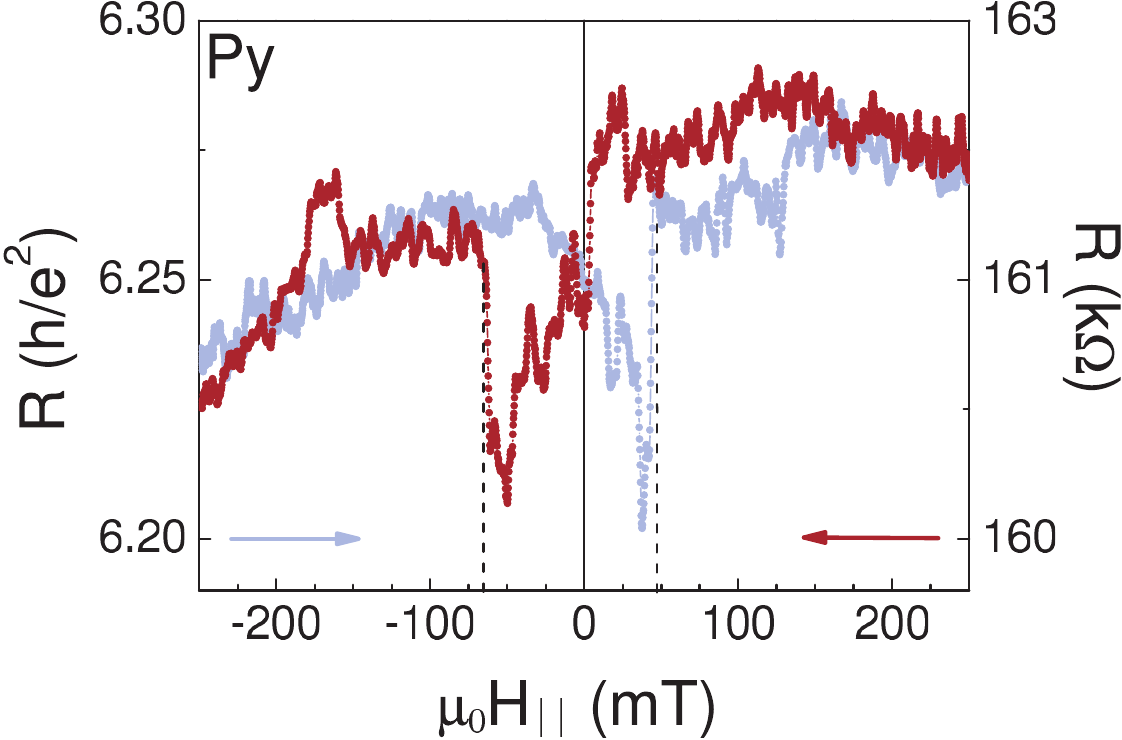}
\caption{\label{TMRMWCNTnegativ}
Differential two-terminal resistance $R=dV/dI$ of a MWCNT pseudo spin valve with
Py electrodes, as a function of an in-plane magnetic field $B_{||}$ along the
electrode long axis. The two electrodes measure $300\un{nm} \times 16\,\mu\text{m}$
and $600\un{nm} \times 16\,\mu\text{m}$ with a thickness of $45\un{nm}$.
The in-plane magnetic field was aligned along the length of the
electrodes. The data shown was measured at $V_\text{g}=350\un{mV}$,
$V_\text{sd}=100\un{mV}$ and  $T=4.2\un{K}$. For each upsweep (in the interval
$[11\un{mT}, 50\un{mT}]$) and downsweep (in the interval
$[-11\un{mT},-69\un{mT}]$), a negative TMR feature of $-1\%$ appears.
}
\end{figure}
An example of the two-terminal resistance
of a MWCNT pseudo spin valve with Py electrodes is shown in Fig.~\ref{TMRMWCNTnegativ}. The differential resistance is plotted as a function of magnetic field
parallel to the length of the electrodes. The measurement displays  negative  TMR
features  of about $-1\%$ for both up- and downsweep within the
intervals $\rm{[-11~mT,~-69~mT]}$ and $\rm{[11~mT,~50~mT]}$. As observed in many of our Py or
$\rm{Pd_{60}Fe_{40}/Fe}$ samples, the measurement suffers from a high
noise level. Unlike the clear TMR feature of Fig.~\ref{TMRMWCNTnegativ},
in many devices random telegraph noise was observed, where
the resistance switched sharply between two or more distinct values.
These jumps were irreproducible in subsequent magnetosweeps and
typically higher in magnitude than the TMR features, limiting the
reproducibility of the magnetoresistance curves.

\subsubsection{TMR in a perpendicular magnetic field}

\begin{figure}[ht]
\includegraphics[width=8.5cm, angle=0,
keepaspectratio]{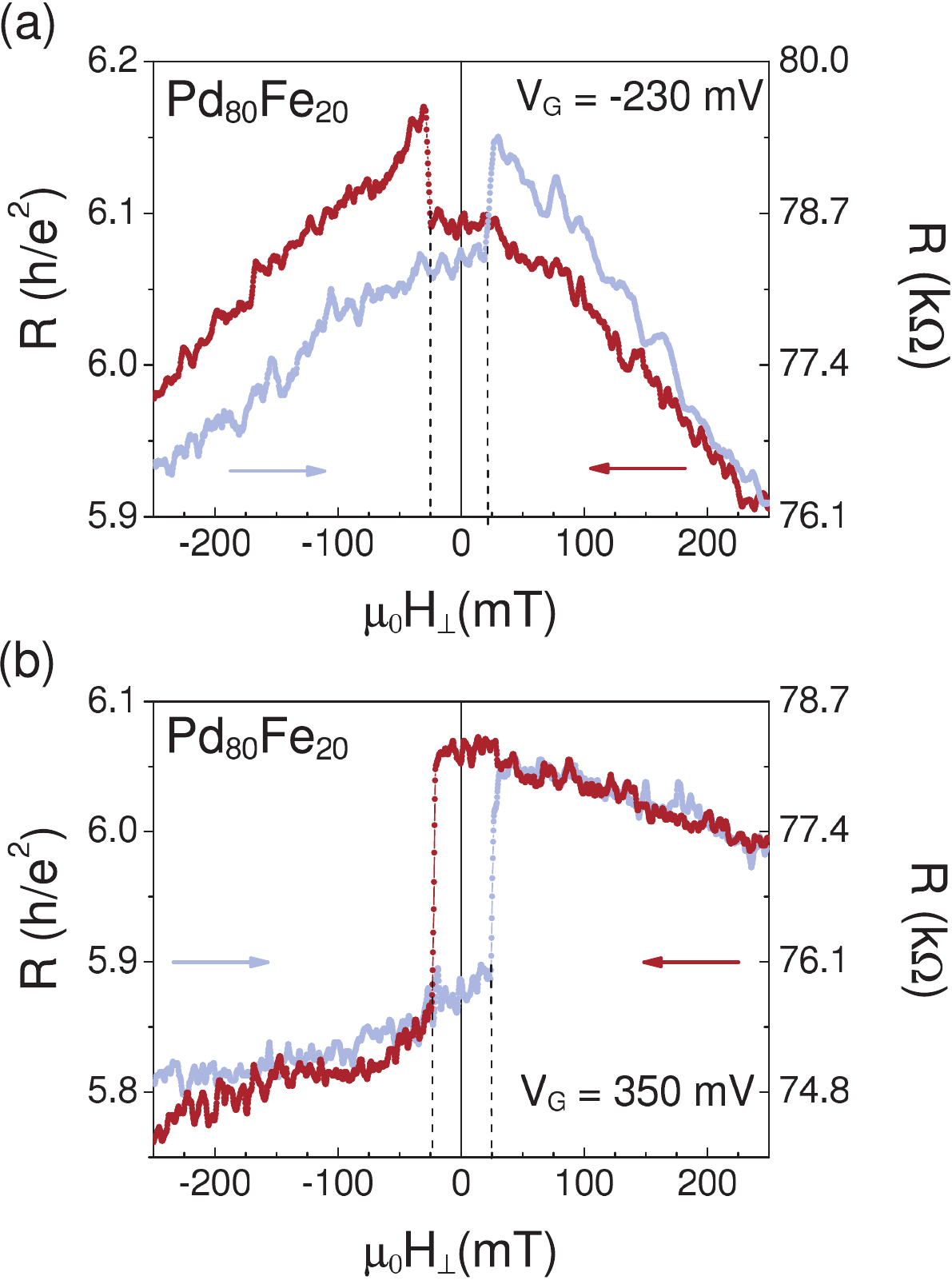}
\caption{
(a)
Differential two-terminal resistance $R=dV/dI$ of a MWCNT pseudo spin valve as a
function of an out-of-plane magnetic field $B_\perp$, applying an ac bias of
$\rm{V_{sd}=100~mV}$ and a gate voltage of $V_g=-230\un{mV}$ at $\rm{4.2~K}$. To access a wider range of energy levels, the pseudo spin valve was patterned on a strongly coupled Al backgate, with the about $\rm{13~nm}$-thick native oxide serving as dielectric.
The two $\rm{Pd_{80}Fe_{20}}$ electrodes measure $\rm{300~nm \times 16 \mu m}$
and $\rm{600~nm \times 16 \mu m}$, with a thickness of $\rm{45~nm}$.
A  resistance jump of 1.2\% occurs in downsweep (in the interval
$\rm{[-23~mT,~-69~mT]}$) and upsweep (in the interval $\rm{[20~mT,~68~mT]}$).
The backswitching from antiparallel to parallel magnetisation configuration is gradual rather than pronouncedly step-like.
(b)
Differential two-terminal resistance $R=dV/dI$ of the same sample as measured in
(a) at a gate voltage of $V_g=350\un{mV}$. At $\pm 25\un{mT}$ for down- and
upsweep respectively, a single magnetoresistance step of $3.1\%$ occurs.
}\label{TMRMWCNT1switchPdFe}
\end{figure}
Two exemplary magnetoresistance sweeps of a MWCNT pseudo spin valve as a
function of an out-of-plane magnetic field are
presented in Fig.\,\ref{TMRMWCNT1switchPdFe}. Also here an out-of-plane field
is found to cause significant magnetoresistance effects.
Fig.~\ref{TMRMWCNT1switchPdFe}(a), recorded at $V_g=-230\un{mV}$, displays a
hysteretical behaviour in the sense that at $\pm23\un{mT}$ there is a sharp
increase in resistance for  up- and downsweep. The return from this
positive TMR feature in antiparallel  electrode magnetisation alignment down to
the resistance in parallel alignment is gradual indicating a gradual shift of
domain walls rather than sharp switching.

In contrast, Fig.\,\ref{TMRMWCNT1switchPdFe}(b), recorded at $V_g=+350\un{mV}$,
exhibits only a single switching event per sweep at $\pm25\un{mT}$. Up- and downsweep are still hysteretical but no longer symmetric with respect to the $B=0$ axis. The resistance increases for the upsweep and decreases for the downsweep between two distinct
resistance levels. Although it occurs at roughly the same magnetic field value as the
first switching event in (a), this behaviour can not be explained with a change
in resistance due to a simple realignment of two ferromagnetic electrodes from parallel to antiparallel magnetisation. The symmetric
occurrence at $\pm 25\un{mT}$  hints that magnetic
switching may stem from only one of the ferromagnetic electrodes whereas in the other electrode, the magnetisation is pinned probably by lattice defects, e.g. grain boundaries. So far the origin of the apparent gate dependence in Fig.\,\ref{TMRMWCNT1switchPdFe} can not be accounted for. It may be a specific feature of MWCNTs since it was not observed yet in SWCNTs.\cite{SWCNTMRfabryPerotMorpurgo, SWCNTMRVgPdNiKontos}

\section{Conclusions}

In conclusion, we have investigated various aspects of carbon nanotube-based
pseudo spin valves with diluted ferromagnetic contacts.

Ferromagnetic contact electrodes of different shapes and materials were
studied for their suitability as
contacts for carbon-nanotube based pseudo spin valves using TEM Lorentz microscopy. This allowed the
identification of materials
and shapes that are in a magnetic single domain state and whose magnetisation
switches at a sharp, reproducible and shape-tunable coercive field.

Permalloy structures displayed the required single-domain
magnetisation reversal. In comparison, $\rm{PdFe}$ alloy strips switched their
magnetisation via multi-domain configurations. This can account for the comparatively lower reproducibility in magnetisation reversal. Furthermore, cross-tie
domain walls were observed as indication of a non-negligible out-of-plane
magnetisation. Multiple fine ripple domains appeared in magnetic double layer structures made from $\rm{PdFe/Fe}$.

For all electrode shapes
investigated, it holds that a larger width lowers the coercive field. End domains were
identified to act as seeds for magnetisation reversal and thereby to decrease the coercive
field. Out of the $\rm{Pd_{60}Fe_{40}/Fe}$ electrode shapes
investigated here, rectangular strips exhibited the most reliable magnetisation
reversal, although a more pronounced end domain structure was observed compared
to pointy and circular end shapes. Large-area ferromagnetic appendices to
electrode strips, as in the case of attached disks or supply lines, were found to
strongly impact the magnetic switching, and to lead to unpredictable behaviour.

Regarding the contact resistance to nanotubes, the properties of PdFe alloys
were found to be significantly better than Py. Device resistances still varied strongly even along the same nanotube.
SQUID measurements on large ensembles of same-sized electrode
structures confirmed the coercive field scale.

Consistent with previous findings  both positive and negative TMR
features\cite{SWCNTMRfabryPerotMorpurgo, SWCNTMRVgPdNiKontos} were observed.
Using PdFe alloy electrodes of $45\un{nm}$ thickness, we were able to observe
TMR for both in-plane and out-of-plane magnetic field direction, confirming the findings from Lorentz microscopy that the out-of-plane magnetisation is non-negligible in such contacts.

The giant paramagnet Palladium seems the obvious base material for carbon nanotube-based spintronics devices. Yet, the thin film magnetic properties of different ferromagnet-Pd alloys can vary strongly, in particular concerning the
out-of-plane magnetisation, and require further optimisation.

We would like to thank H. S. J. van der Zant, A. Morpurgo, and B. Witkamp for
helping us set up our CVD system, L. Forro and C. Miko for providing the multi-walled
carbon nanotubes, and M. Aprili for fruitful discussion. Thanks also go to E. Lipp for proofreading. This work was supported by the DFG
within the SFB 689 and the EU FP6 CARDEQ project.


\begin{thebibliography}{24}
\expandafter\ifx\csname natexlab\endcsname\relax\def\natexlab#1{#1}\fi
\expandafter\ifx\csname bibnamefont\endcsname\relax
  \def\bibnamefont#1{#1}\fi
\expandafter\ifx\csname bibfnamefont\endcsname\relax
  \def\bibfnamefont#1{#1}\fi
\expandafter\ifx\csname citenamefont\endcsname\relax
  \def\citenamefont#1{#1}\fi
\expandafter\ifx\csname url\endcsname\relax
  \def\url#1{\texttt{#1}}\fi
\expandafter\ifx\csname urlprefix\endcsname\relax\def\urlprefix{URL }\fi
\providecommand{\bibinfo}[2]{#2}
\providecommand{\eprint}[2][]{\url{#2}}

\bibitem[{\citenamefont{Prinz}(1998)}]{citeulike:1119796}
\bibinfo{author}{\bibfnamefont{G.~A.} \bibnamefont{Prinz}},
  \bibinfo{journal}{Science} \textbf{\bibinfo{volume}{282}},
  \bibinfo{pages}{1660} (\bibinfo{year}{1998}).

\bibitem[{\citenamefont{Cottet et~al.}(2006)\citenamefont{Cottet, Kontos,
  Sahoo, Man, Choi, Belzig, Bruder, Morpurgo, and
  Sch\"onenberger}}]{NanospintronicsCNTferroKontosREVIEW}
\bibinfo{author}{\bibfnamefont{A.}~\bibnamefont{Cottet}},
  \bibinfo{author}{\bibfnamefont{T.}~\bibnamefont{Kontos}},
  \bibinfo{author}{\bibfnamefont{S.}~\bibnamefont{Sahoo}},
  \bibinfo{author}{\bibfnamefont{H.~T.} \bibnamefont{Man}},
  \bibinfo{author}{\bibfnamefont{M.~S.} \bibnamefont{Choi}},
  \bibinfo{author}{\bibfnamefont{W.}~\bibnamefont{Belzig}},
  \bibinfo{author}{\bibfnamefont{C.}~\bibnamefont{Bruder}},
  \bibinfo{author}{\bibfnamefont{A.~F.} \bibnamefont{Morpurgo}},
  \bibnamefont{and}
  \bibinfo{author}{\bibfnamefont{C.}~\bibnamefont{Sch\"onenberger}},
  \bibinfo{journal}{Semic. Sci. and Techn.} \textbf{\bibinfo{volume}{21}},
  \bibinfo{pages}{S78} (\bibinfo{year}{2006}).

\bibitem[{\citenamefont{Hauptmann et~al.}(2008)\citenamefont{Hauptmann, Paaske,
  and Lindelof}}]{KondoNiSWCNTLinedelof}
\bibinfo{author}{\bibfnamefont{J.~R.} \bibnamefont{Hauptmann}},
  \bibinfo{author}{\bibfnamefont{J.}~\bibnamefont{Paaske}}, \bibnamefont{and}
  \bibinfo{author}{\bibfnamefont{P.~E.} \bibnamefont{Lindelof}},
  \bibinfo{journal}{Nat. Phys.} \textbf{\bibinfo{volume}{4}},
  \bibinfo{pages}{373} (\bibinfo{year}{2008}).

\bibitem[{\citenamefont{Buitelaar et~al.}(2002)\citenamefont{Buitelaar,
  Bachtold, Nussbaumer, Iqbal, and
  Sch\"{o}nenberger}}]{mwcntkondoSchoenenberger}
\bibinfo{author}{\bibfnamefont{M.~R.} \bibnamefont{Buitelaar}},
  \bibinfo{author}{\bibfnamefont{A.}~\bibnamefont{Bachtold}},
  \bibinfo{author}{\bibfnamefont{T.}~\bibnamefont{Nussbaumer}},
  \bibinfo{author}{\bibfnamefont{M.}~\bibnamefont{Iqbal}}, \bibnamefont{and}
  \bibinfo{author}{\bibfnamefont{C.}~\bibnamefont{Sch\"{o}nenberger}},
  \bibinfo{journal}{Phys. Rev. Lett.} \textbf{\bibinfo{volume}{88}},
  \bibinfo{pages}{156801} (\bibinfo{year}{2002}).

\bibitem[{\citenamefont{Man et~al.}(2006)\citenamefont{Man, Wever, and
  Morpurgo}}]{SWCNTMRfabryPerotMorpurgo}
\bibinfo{author}{\bibfnamefont{H.~T.} \bibnamefont{Man}},
  \bibinfo{author}{\bibfnamefont{I.~J.~W.} \bibnamefont{Wever}},
  \bibnamefont{and} \bibinfo{author}{\bibfnamefont{A.~F.}
  \bibnamefont{Morpurgo}}, \bibinfo{journal}{Phys. Rev. B}
  \textbf{\bibinfo{volume}{73}} (\bibinfo{year}{2006}).

\bibitem[{\citenamefont{Javey et~al.}(2003{\natexlab{a}})\citenamefont{Javey,
  Guo, Wang, Lundstrom, and Dai}}]{PdswcntballistictransparentContactDai}
\bibinfo{author}{\bibfnamefont{A.}~\bibnamefont{Javey}},
  \bibinfo{author}{\bibfnamefont{J.}~\bibnamefont{Guo}},
  \bibinfo{author}{\bibfnamefont{Q.}~\bibnamefont{Wang}},
  \bibinfo{author}{\bibfnamefont{M.}~\bibnamefont{Lundstrom}},
  \bibnamefont{and} \bibinfo{author}{\bibfnamefont{H.}~\bibnamefont{Dai}},
  \bibinfo{journal}{Nature} \textbf{\bibinfo{volume}{424}},
  \bibinfo{pages}{654} (\bibinfo{year}{2003}{\natexlab{a}}).

\bibitem[{\citenamefont{Stirling et~al.}(1972)\citenamefont{Stirling, Cowley,
  and Stringfellow}}]{PdFegiantparamagnetic}
\bibinfo{author}{\bibfnamefont{W.~G.} \bibnamefont{Stirling}},
  \bibinfo{author}{\bibfnamefont{R.~A.} \bibnamefont{Cowley}},
  \bibnamefont{and} \bibinfo{author}{\bibfnamefont{M.~W.}
  \bibnamefont{Stringfellow}}, \bibinfo{journal}{J. Phys. F: Metal Physics}
  \textbf{\bibinfo{volume}{2}}, \bibinfo{pages}{421} (\bibinfo{year}{1972}).

\bibitem[{\citenamefont{Sahoo et~al.}(2005{\natexlab{a}})\citenamefont{Sahoo,
  Kontos, Sch\"onenberger, and S\"urgers}}]{MWCNTspinInjTranspKontos}
\bibinfo{author}{\bibfnamefont{S.}~\bibnamefont{Sahoo}},
  \bibinfo{author}{\bibfnamefont{T.}~\bibnamefont{Kontos}},
  \bibinfo{author}{\bibfnamefont{C.}~\bibnamefont{Sch\"onenberger}},
  \bibnamefont{and}
  \bibinfo{author}{\bibfnamefont{C.}~\bibnamefont{S\"urgers}},
  \bibinfo{journal}{Appl. Phys. Lett.} \textbf{\bibinfo{volume}{86}},
  \bibinfo{pages}{112109} (\bibinfo{year}{2005}{\natexlab{a}}).

\bibitem[{\citenamefont{Zhao et~al.}(2002)\citenamefont{Zhao, M\"onch, M\"uhl,
  Vinzelberg, and Schneider}}]{mwcntmagnetotransportMagneticEdge-domain}
\bibinfo{author}{\bibfnamefont{B.}~\bibnamefont{Zhao}},
  \bibinfo{author}{\bibfnamefont{I.}~\bibnamefont{M\"onch}},
  \bibinfo{author}{\bibfnamefont{T.}~\bibnamefont{M\"uhl}},
  \bibinfo{author}{\bibfnamefont{H.}~\bibnamefont{Vinzelberg}},
  \bibnamefont{and} \bibinfo{author}{\bibfnamefont{C.~M.}
  \bibnamefont{Schneider}}, \bibinfo{journal}{J. Appl. Phys.}
  \textbf{\bibinfo{volume}{91}}, \bibinfo{pages}{7026} (\bibinfo{year}{2002}).

\bibitem[{\citenamefont{Sahoo et~al.}(2005{\natexlab{b}})\citenamefont{Sahoo,
  Kontos, Furer, Hoffmann, Graber, Cottet, and
  Sch\"onenberger}}]{SWCNTMRVgPdNiKontos}
\bibinfo{author}{\bibfnamefont{S.}~\bibnamefont{Sahoo}},
  \bibinfo{author}{\bibfnamefont{T.}~\bibnamefont{Kontos}},
  \bibinfo{author}{\bibfnamefont{J.}~\bibnamefont{Furer}},
  \bibinfo{author}{\bibfnamefont{C.}~\bibnamefont{Hoffmann}},
  \bibinfo{author}{\bibfnamefont{M.}~\bibnamefont{Graber}},
  \bibinfo{author}{\bibfnamefont{A.}~\bibnamefont{Cottet}}, \bibnamefont{and}
  \bibinfo{author}{\bibfnamefont{C.}~\bibnamefont{Sch\"onenberger}},
  \bibinfo{journal}{Nat. Phys.} \textbf{\bibinfo{volume}{1}},
  \bibinfo{pages}{99} (\bibinfo{year}{2005}{\natexlab{b}}).

\bibitem[{\citenamefont{Julliere}(1975)}]{Julliere1975225}
\bibinfo{author}{\bibfnamefont{M.}~\bibnamefont{Julliere}},
  \bibinfo{journal}{Physics Letters A} \textbf{\bibinfo{volume}{54}},
  \bibinfo{pages}{225 } (\bibinfo{year}{1975}).

\bibitem[{\citenamefont{Kim et~al.}(2002)\citenamefont{Kim, So, Kim, and
  Kim}}]{MRSWCNTCoKim}
\bibinfo{author}{\bibfnamefont{J.-R.} \bibnamefont{Kim}},
  \bibinfo{author}{\bibfnamefont{H.~M.} \bibnamefont{So}},
  \bibinfo{author}{\bibfnamefont{J.-J.} \bibnamefont{Kim}}, \bibnamefont{and}
  \bibinfo{author}{\bibfnamefont{J.}~\bibnamefont{Kim}},
  \bibinfo{journal}{Phys. Rev. B} \textbf{\bibinfo{volume}{66}},
  \bibinfo{pages}{233401} (\bibinfo{year}{2002}).

\bibitem[{\citenamefont{Chapman}(1999)}]{LoMicroReview}
\bibinfo{author}{\bibfnamefont{J.}~\bibnamefont{Chapman}},
  \bibinfo{journal}{Journal of Magnetism and Magnetic Materials}
  \textbf{\bibinfo{volume}{200}}, \bibinfo{pages}{729} (\bibinfo{year}{1999}).

\bibitem[{\citenamefont{Lim}(2002)}]{LorentzMicroSpinValveChapman}
\bibinfo{author}{\bibfnamefont{C.}~\bibnamefont{Lim}},
  \bibinfo{journal}{Journal of Magnetism and Magnetic Materials}
  \textbf{\bibinfo{volume}{238}}, \bibinfo{pages}{301} (\bibinfo{year}{2002}).

\bibitem[{\citenamefont{Uhlig and
  Zweck}(2004)}]{LorentzMicroPyCircSwitchingZweck}
\bibinfo{author}{\bibfnamefont{T.}~\bibnamefont{Uhlig}} \bibnamefont{and}
  \bibinfo{author}{\bibfnamefont{J.}~\bibnamefont{Zweck}},
  \bibinfo{journal}{Phys. Rev. Lett.} \textbf{\bibinfo{volume}{93}},
  \bibinfo{pages}{047203} (\bibinfo{year}{2004}).

\bibitem[{\citenamefont{Hubert and Sch\"{a}fer}(1998)}]{MagneticDomains}
\bibinfo{author}{\bibfnamefont{A.}~\bibnamefont{Hubert}} \bibnamefont{and}
  \bibinfo{author}{\bibfnamefont{R.}~\bibnamefont{Sch\"{a}fer}},
  \emph{\bibinfo{title}{Magnetic Domains}} (\bibinfo{publisher}{Springer},
  \bibinfo{year}{1998}), \bibinfo{edition}{1st} ed.

\bibitem[{\citenamefont{Gomez et~al.}(1999)\citenamefont{Gomez, Luu, Pak, Kirk,
  and Chapman}}]{Pydomainconfig}
\bibinfo{author}{\bibfnamefont{R.~D.} \bibnamefont{Gomez}},
  \bibinfo{author}{\bibfnamefont{T.~V.} \bibnamefont{Luu}},
  \bibinfo{author}{\bibfnamefont{A.~O.} \bibnamefont{Pak}},
  \bibinfo{author}{\bibfnamefont{K.~J.} \bibnamefont{Kirk}}, \bibnamefont{and}
  \bibinfo{author}{\bibfnamefont{J.~N.} \bibnamefont{Chapman}},
  \bibinfo{journal}{J. Appl. Phys.} \textbf{\bibinfo{volume}{85}},
  \bibinfo{pages}{6163} (\bibinfo{year}{1999}).

\bibitem[{\citenamefont{Shalyguina et~al.}(2002)\citenamefont{Shalyguina, Shin,
  and Abrosimova}}]{PyStripSizeDepMagn}
\bibinfo{author}{\bibfnamefont{E.~E.} \bibnamefont{Shalyguina}},
  \bibinfo{author}{\bibfnamefont{K.~H.} \bibnamefont{Shin}}, \bibnamefont{and}
  \bibinfo{author}{\bibfnamefont{N.~M.} \bibnamefont{Abrosimova}},
  \bibinfo{journal}{Journal of Magnetism and Magnetic Materials}
  \textbf{\bibinfo{volume}{239}}, \bibinfo{pages}{252} (\bibinfo{year}{2002}).

\bibitem[{\citenamefont{Yu et~al.}(1999)\citenamefont{Yu, R\"{u}diger, Thomas,
  Parkin, and Kent}}]{Femagnetic-switchingneedlerectangle}
\bibinfo{author}{\bibfnamefont{J.}~\bibnamefont{Yu}},
  \bibinfo{author}{\bibfnamefont{U.}~\bibnamefont{R\"{u}diger}},
  \bibinfo{author}{\bibfnamefont{L.}~\bibnamefont{Thomas}},
  \bibinfo{author}{\bibfnamefont{S.~S.~P.} \bibnamefont{Parkin}},
  \bibnamefont{and} \bibinfo{author}{\bibfnamefont{A.~D.} \bibnamefont{Kent}},
  \bibinfo{journal}{J. Appl. Phys.} \textbf{\bibinfo{volume}{85}},
  \bibinfo{pages}{5501} (\bibinfo{year}{1999}).

\bibitem[{\citenamefont{Kirk et~al.}(1999)\citenamefont{Kirk, Chapman, and
  Wilkinson}}]{LoMicroPyNeedlesRectangles}
\bibinfo{author}{\bibfnamefont{K.~J.} \bibnamefont{Kirk}},
  \bibinfo{author}{\bibfnamefont{J.~N.} \bibnamefont{Chapman}},
  \bibnamefont{and} \bibinfo{author}{\bibfnamefont{C.~D.~W.}
  \bibnamefont{Wilkinson}}, \bibinfo{journal}{J. Appl. Phys.}
  \textbf{\bibinfo{volume}{85}}, \bibinfo{pages}{5237} (\bibinfo{year}{1999}).

\bibitem[{\citenamefont{Huber}(2005)}]{michaelhuber}
\bibinfo{author}{\bibfnamefont{M.}~\bibnamefont{Huber}} (\bibinfo{year}{2005}),
  \bibinfo{note}{unpublished work}.

\bibitem[{\citenamefont{McCartney and
  Smith}(1997)}]{FeLorentzmicroscopyRippledomain}
\bibinfo{author}{\bibfnamefont{M.~R.} \bibnamefont{McCartney}}
  \bibnamefont{and} \bibinfo{author}{\bibfnamefont{D.~J.} \bibnamefont{Smith}},
  \bibinfo{journal}{Scanning Microscopy} \textbf{\bibinfo{volume}{11}},
  \bibinfo{pages}{335} (\bibinfo{year}{1997}).

\bibitem[{\citenamefont{Kong et~al.}(1998)\citenamefont{Kong, Soh, Cassell,
  Quate, and Dai}}]{growth}
\bibinfo{author}{\bibfnamefont{J.}~\bibnamefont{Kong}},
  \bibinfo{author}{\bibfnamefont{H.~T.} \bibnamefont{Soh}},
  \bibinfo{author}{\bibfnamefont{A.~M.} \bibnamefont{Cassell}},
  \bibinfo{author}{\bibfnamefont{C.~F.} \bibnamefont{Quate}}, \bibnamefont{and}
  \bibinfo{author}{\bibfnamefont{H.}~\bibnamefont{Dai}},
  \bibinfo{journal}{Nature} \textbf{\bibinfo{volume}{395}},
  \bibinfo{pages}{878} (\bibinfo{year}{1998}).

\bibitem[{\citenamefont{Javey et~al.}(2003{\natexlab{b}})\citenamefont{Javey,
  Guo, Wang, Lundstrom, and Dai}}]{FETcntPdDai}
\bibinfo{author}{\bibfnamefont{A.}~\bibnamefont{Javey}},
  \bibinfo{author}{\bibfnamefont{J.}~\bibnamefont{Guo}},
  \bibinfo{author}{\bibfnamefont{Q.}~\bibnamefont{Wang}},
  \bibinfo{author}{\bibfnamefont{M.}~\bibnamefont{Lundstrom}},
  \bibnamefont{and} \bibinfo{author}{\bibfnamefont{H.}~\bibnamefont{Dai}},
  \bibinfo{journal}{Nature} \textbf{\bibinfo{volume}{424}},
  \bibinfo{pages}{654} (\bibinfo{year}{2003}{\natexlab{b}}).

\end{thebibliography}
\end{document}